\documentclass[11pt,oneside,english,reqno]{amsart}
\usepackage{mathpazo}
\usepackage[scaled=0.92]{helvet}
\usepackage{courier}
\usepackage[T1]{fontenc}
\usepackage[latin9]{inputenc}
\usepackage[a4paper]{geometry}
\usepackage{color}
\usepackage{float}
\usepackage{units}
\usepackage{amsbsy}
\usepackage{amstext}
\usepackage{amsthm}
\usepackage{amssymb}
\usepackage{graphicx}
\usepackage{setspace}
\usepackage[authoryear]{natbib}
\PassOptionsToPackage{normalem}{ulem}
\usepackage{ulem}
\makeatletter

\floatstyle{ruled}
\newfloat{algorithm}{tbp}{loa}
\providecommand{\algorithmname}{Algorithm}
\floatname{algorithm}{\protect\algorithmname}

\numberwithin{equation}{section}
\numberwithin{figure}{section}

\theoremstyle{plain}

\numberwithin{equation}{section}
\usepackage{algorithm,algpseudocode}
\usepackage{color}

\makeatother

\usepackage{babel}
\begin{document}

\title[Efficient covariance approximations for large sparse precision matrices]{Efficient covariance approximations for large sparse precision matrices}

\author{Per Sidén, Finn Lindgren, David Bolin and Mattias Villani}

\thanks{Per Sidén: \textit{Division of Statistics and Machine Learning, Dept.
of Computer and Information Science, Linköping University, SE-581
83 Linköping, Sweden}. \textit{E-mail: per.siden@liu.se. Corresponding
author.} \\
Finn Lindgren: \textit{School of Mathematics, The University of Edinburgh,
James Clerk Maxwell Building, The King's Building, Peter Guthrie Tait
Road, Edinburgh, EH9 3FD, United Kingdom. E-mail: finn.lindgren@ed.ac.uk.}\\
David Bolin: \textit{Division of Mathematical Statistics, Dept. of
Mathematical Sciences, Chalmers and University of Gothenburg, SE-412
96 Göteborg, Sweden. E-mail: david.bolin@chalmers.se.}\\
Mattias Villani: \textit{Division of Statistics and Machine Learning,
Dept. of Computer and Information Science, Linköping University, SE-581
83 Linköping, Sweden}. \textit{E-mail: mattias.villani@liu.se}.}
\begin{abstract}
The use of sparse precision (inverse covariance) matrices has become
popular because they allow for efficient algorithms for joint inference
in high-dimensional models. Many applications require the computation
of certain elements of the covariance matrix, such as the marginal
variances, which may be non-trivial to obtain when the dimension is
large. This paper introduces a fast Rao-Blackwellized Monte Carlo
sampling-based method for efficiently approximating selected elements
of the covariance matrix. The variance and confidence bounds of the
approximations can be precisely estimated without additional computational
costs. Furthermore, a method that iterates over subdomains is introduced,
and is shown to additionally reduce the approximation errors to practically
negligible levels in an application on functional magnetic resonance
imaging data. Both methods have low memory requirements, which is
typically the bottleneck for competing direct methods.
\end{abstract}

\maketitle

\section{Introduction}

We consider the problem of computing selected elements of the covariance
matrix $\boldsymbol{\Sigma}$ of a multivariate normal distribution,
which is parameterized using the precision matrix $\mathbf{Q}=\boldsymbol{\Sigma}^{-1}$.
Specifying models with the precision matrix rather than the covariance
matrix is useful (or even necessary) in many high-dimensional applications,
since it allows for a sparse representation, typically leading to
smaller time and memory complexity. Their use has a long history in
spatial statistics \citep{Besag1974,Isham2004}, image processing
\citep{Jeng1991}, and probabilistic graphical models \citep{Lauritzen1996,Malioutov2006}.

The desire to compute certain elements of the covariance matrix arises
in many applications. If $\mathbf{Q}$ is the posterior precision
matrix in a Bayesian analysis, then the diagonal of $\boldsymbol{\Sigma}$
contains the posterior marginal variances, which are often presented
as a measure of marginal uncertainty. Furthermore, joint posterior
statistics of larger subdomains will normally also require the computation
of certain off-diagonal elements of $\boldsymbol{\Sigma}$. This is
for example the case when computing the posterior probability of exceeding
a threshold in a specific subdomain, which is the topic of \citet{Bolin2014a},
and which has applications in temperature modeling \citep{Furrer2007},
astrophysics \citep{beaky1992topology} and brain imaging \citep{Sid??n2017}.
Computing submatrices of $\boldsymbol{\Sigma}$ is also needed for
characterizing the uncertainty of a robot's location in an unknown
environment \citep{Thrun2005}. 

Even though $\mathbf{Q}$ is sparse, $\boldsymbol{\Sigma}$ is dense
in general, and the naive direct inversion $\boldsymbol{\Sigma}=\mathbf{Q}^{-1}$
is not an option even for relatively small dimensions. Fortunately,
as described in the next section, when only selected elements of $\boldsymbol{\Sigma}$
are required, a number of less computationally intensive exact methods
exist in the literature. However, for modern applications the dimensionality
of the problem can be too large even for these methods to be computationally
feasible. Usually the bottleneck is in the memory requirements, even
though computation times can also be unpleasantly long. This often
leads to investigators choosing to perform their analyses on smaller
subsets of the data independently or at a lower resolution than desired.
Another situation in which the memory is normally a limitation is
when performing these operations on robots or other embedded systems.

In this paper, we develop a fast Rao-Blackwellized Monte Carlo sampling-based
method for approximating the elements of the covariance matrix, and
show its efficiency compared to existing sampling-based methods. We
further show that the variances and confidence bounds of the approximations
can be cheaply computed by inserting the approximated values into
analytical expressions. In addition, a second, more exact method is
developed, which by using the estimates from the first method as starting
values and by iterating over subdomains, produces estimates with negligible
error in practice. Both methods build on decompositions of the domain
on which the GMRF is defined, into subdomains that can be processed
nearly independently, leading to low memory requirements and algorithms
that are easily parallelized. We evaluate the methods on precision
matrices from theoretical models and on a posterior precision matrix
from a functional magnetic resonance imaging (fMRI) experiment.

The outline of the article follows. In the next section, we give a
theoretical background to the problem and an overview of existing
methods and their limitations. We present the developed methods in
Section~\ref{sec:Methods} and numerically evaluate their performance
in Section~\ref{sec:Results}. Section~\ref{sec:Discussion} is
a discussion and Section~\ref{sec:Conclusions} concludes. A Matlab
implementation of the methods in the article is available at \uline{https://github.com/psiden/CovApprox}.

\section{Background and literature review\label{sec:Background}}

We assume that $\mathbf{x}\sim\mathsf{N}\left(\boldsymbol{\mu},\mathbf{Q}^{-1}\right)$
is an $N$-dimensional multivariate normally distributed random variable,
and that $\mathbf{Q}$ is a sparse symmetric positive definite precision
matrix. Such a distribution is commonly referred to as a Gaussian
Markov random field (GMRF) \citep{Isham2004} and the sparsity pattern
of $\mathbf{Q}$ has a natural interpretation in that an element $Q_{i,j}$
is zero if and only if the corresponding elements $x_{i}$ and $x_{j}$
are conditionally independent given all other elements. We assume,
for simplicity and without loss of generality, that $\boldsymbol{\mu}=\mathbf{0}$.

Selected inversion refers to the computation of $\boldsymbol{\Sigma}_{\mathcal{S}}$
for some set of indices \\
$\mathcal{S}\subset\left\{ \left(i,j\right);1\leq i,j\leq N\right\} $,
with $\left|\mathcal{S}\right|\ll N^{2}$, for example $\mathcal{S}=\mathcal{S}_{I}=\left\{ \left(i,j\right);i=j\right\} $
gives the diagonal. We will also use the (slightly abusive) notation
$\boldsymbol{\sigma}^{\mathbf{2}}=\left[\sigma_{1}^{2},\ldots,\sigma_{N}^{2}\right]=\boldsymbol{\Sigma}_{\mathcal{S}_{I}},$
to denote the marginal variances. Other commonly appearing examples
of index sets for selected inversion are $\mathcal{S}_{\mathbf{aa}^{T}}=\left\{ \left(i,j\right);a_{i}\neq0,a_{j}\neq0\right\} $,
for some sparse column vector $\mathbf{a}$, and $\mathcal{S}_{\mathbf{R}}=\left\{ \left(i,j\right);R_{ij}\neq0\right\} $,
for some sparse symmetric matrix $\mathbf{R}$. The first can be used
to compute $Var\left(\mathbf{a}^{T}\text{\textbf{x}}\right)=\mathbf{a}^{T}\boldsymbol{\Sigma}\mathbf{a}=\sum_{i,j}a_{i}a_{j}\Sigma_{i,j}$,
and the second when computing $tr\left(\mathbf{R}\boldsymbol{\Sigma}\right)=\sum_{i,j}R_{i,j}\Sigma_{i,j}$,
which is commonly needed in some inference methods such as the expectation
maximization (EM) algorithm \citep{Bolin2009} and variational Bayes
(VB) \citep{Solis-Trapala2009}. Depending on $\mathcal{S}$ and the
sparsity pattern of $\mathbf{Q}$, different methods for selected
inversion might be preferable.

A naive method for selected inversion that always works in theory
for any $\mathcal{S}$, is of course to completely compute $\boldsymbol{\Sigma}=\mathbf{Q}^{-1}$
using a standard method, for example Gaussian elimination, and then
extract $\boldsymbol{\Sigma}_{\mathcal{S}}$. Such a method is of
time complexity $O\left(N^{3}\right)$ and memory complexity $O$$\left(N^{2}\right)$
which is prohibitive even for rather small values of $N$. By exploiting
the sparsity patterns in $\mathcal{S}$ and $\mathbf{Q}$, this complexity
can often be greatly reduced. 

Another trivial idea for selected inversion is that column $j$ of
$\boldsymbol{\Sigma}$ can be computed by solving $\mathbf{Qz}=\mathbf{e}_{j}$
for $\mathbf{z}$, where $\mathbf{e}_{j}$ is the $j$th column of
the $N\times N$ identity matrix. The computational cost of this operation
may be high for large $N$, but can be greatly reduced by using iterative
methods such as the preconditioned conjugate gradient (PCG) algorithm
\citep{Manteuffel1980,barrett1994templates}. This method produces
an approximate solution by iteratively minimizing the relative residual
$\left\Vert \mathbf{Qz}-\mathbf{e}_{j}\right\Vert /\left\Vert \mathbf{e}_{j}\right\Vert $
until it decreases below some specified level $\delta$, that can
be set arbitrarily low. The time complexity of iterative methods can
be nearly linear in $N$ for diagonally dominant matrices $\mathbf{Q}$,
which are matrices $Q_{i,i}>\sum_{i\neq j}\left|Q_{i,j}\right|$ for
all $i$ \citep{Spielman2004}. However, in general PCG has complexity
$O\left(m\sqrt{\kappa}\right)$, where $m$ is the number of nonzero
elements in $\mathbf{Q}$ and $\kappa$ is its condition number. This
for example gives complexity $O\left(N^{1+1/d}\right)$ if $\mathbf{Q}$
is obtained from a finite element approximation of a second-order
elliptic boundary value problem posed on a $d$-dimensional domain
\citep{Shewchuk1994}. This strategy will therefore have at least
quadratic complexity when the selected elements are in all columns
(for example when computing $\boldsymbol{\Sigma}_{\mathcal{S}_{I}}$),
which will often be too costly, but it can be useful when the number
of selected elements is small.

Direct methods for selected inversion usually rely on first computing
the Cholesky decomposition $\mathbf{LL}^{T}=\mathbf{Q}$, where $\mathbf{L}$
is lower triangular. This operation also takes $O\left(N^{3}\right)$
time in general, but by using reordering techniques, for example approximate
minimum degree reordering \citep{Amestoy1996}, it can generally be
reduced to $O\left(N^{3/2}\right)$ for 2D problems and $O\left(N^{2}\right)$
for 3D problems \citep{Isham2004}. It is, however, the memory requirements
that normally make these methods unfeasible \citep{Aune2014}. The
complexity mainly depends on the sparsity pattern of $\mathbf{L}$,
whose dependency on $\mathbf{Q}$ is nicely explained from a graph
theoretical point of view in \citet{Vandenberghe2014a}. We denote
the index set of the symbolic Cholesky factorization by $\mathcal{L}_{\mathbf{Q}}$
(edges in the chordal extension of the graph corresponding to $\mathbf{Q}$),
which has the property that $\mathcal{L}_{\mathbf{Q}}\supseteq\mathcal{S}_{\mathbf{L}+\mathbf{L}^{T}}\cup\mathcal{S}_{\mathbf{Q}}$,
but in most cases $\mathcal{L}_{\mathbf{Q}}=\mathcal{S}_{\mathbf{L}+\mathbf{L}^{T}}$.
Largely speaking, the complexity is low whenever the fill-in $\mathcal{L}_{\mathbf{Q}}\setminus\mathcal{S}_{\mathbf{Q}}$
is small.

The probably oldest idea for direct selected inversion, referred to
as the Takahashi equations \citep{Takahashi1973,Erisman1975}, is
nicely presented and compactly derived in \citet{Rue2007}. A statistical
derivation in the same article begins by noting that 
\begin{equation}
x_{i}|\mathbf{x}_{i+1:N}\sim\mathsf{N}\left(-\frac{1}{L_{i,i}}\sum_{k=i+1}^{N}L_{k,i}x_{k},1/L_{i,i}^{2}\right),\label{eq:CholNormalDistr}
\end{equation}
which provides a sequential representation of the GMRF. For $j\geq i$
it is straightforward to derive that
\begin{eqnarray}
\Sigma_{ij} & = & E\left(x_{i}x_{j}\right)=E\left[E\left(x_{i}x_{j}|\mathbf{x}_{i+1:N}\right)\right]=\frac{\mathbb{1}_{\left\{ i=j\right\} }}{L_{i,i}^{2}}-\frac{1}{L_{i,i}}\sum_{k=i+1}^{N}L_{k,i}\Sigma_{k,j}.\label{eq:Takahashi}
\end{eqnarray}
By iterating backwards, for $i=N,\ldots,1$ and for each $i,$ $j=N,\ldots,i$,
one can compute the full $\boldsymbol{\Sigma}$ recursively. Furthermore,
because of the sparsity structure of $\mathbf{L}$, many terms of
the sum in Eq.~(\ref{eq:Takahashi}) will be zero, and the authors
show that it is enough to compute $\Sigma_{ij}$ for iterations where
$\left(i,j\right)\in\mathcal{L}_{\mathbf{Q}}$ and to sum over indices
where $\left(k,i\right)\in\mathcal{L}_{\mathbf{Q}}$ for the computations
to be correct for all of $\boldsymbol{\Sigma}_{\mathcal{L}_{\mathbf{Q}}}$.
Therefore, if the selected elements $\mathcal{S}$ form a subset of
$\mathcal{L}_{\mathbf{Q}}$, which is true for example for $\mathcal{S}_{\mathbf{Q}}$
and $\mathcal{S}_{I}$, this method is sufficient for computing the
selected inverse. If $\mathcal{S}=\mathcal{S}_{\mathbf{R}}$ is not
a subset of $\mathcal{L}_{\mathbf{Q}}$, then one could easily show
that it will be sufficient to iterate over the Takahashi equations
for indices in $\mathcal{L}_{\left|\mathbf{Q}\right|+\left|\mathbf{R}\right|}$
instead, by applying Theorem 1 in \citet{Rue2007} on the graph corresponding
to $\left|\mathbf{Q}\right|+\left|\mathbf{R}\right|$. The time complexity
of solving the Takahashi equations is similar that of Cholesky factorization
as illustrated in \citet[Fig. 9.5]{Vandenberghe2014a}. 

In the literature on numerical linear algebra, some effort has in
recent years been devoted to improving this and similar direct methods
\citep{Li2008,Lin2011a,Lin2011,Rouet2012,Amestoy2012,Vandenberghe2014a,Amestoy2015,Xia2015,Jacquelin2015}.
The new techniques use various reorderings, multifrontal and supernodal
strategies, cleverly adapted to the sparsity structure in $\mathbf{Q}$,
in order to distribute computations and storage in efficient ways,
see \citet{Vandenberghe2014a} for an in depth explanation. For problems
of moderate size, these methods are very competitive, but will always
have memory limitations for problems of higher dimensionality.

In the field of probabilistic graphical models, belief propagation
(BP) and loopy belief propagation (LBP) \citep{Pearl1988,Malioutov2006}
are well-known algorithms for inferring the marginal distributions
of the nodes of a graphical model, by iteratively passing messages
between neighboring nodes using various message passing schemes. When
applied to GMRFs, these algorithms compute the means and marginal
variances of all nodes, resulting in the covariance matrix diagonal.
In this case, each message passing step is related to computing the
Schur complement $\boldsymbol{\Sigma}_{\mathcal{I},\mathcal{I}}=\left(\mathbf{Q}_{\mathcal{I},\mathcal{I}}-\mathbf{Q}_{\mathcal{I},\mathcal{I}^{c}}\mathbf{Q}_{\mathcal{I}^{c},\mathcal{I}^{c}}^{-1}\mathbf{Q}_{\mathcal{I}^{c},\mathcal{I}}\right)^{-1}$,
applied to a single node, $\mathcal{I}=\left\{ i\right\} $, see \citet{Malioutov2006}
for details. Of course, $\mathbf{Q}_{\mathcal{I}^{c},\mathcal{I}^{c}}^{-1}$
is unavailable for large graphs, but an approximation, based on saved
results from previous iterations of the algorithm, can be computed
for elements corresponding to neighbors of node $i$. These are the
only ones required due to the sparsity pattern of $\mathbf{Q}_{\mathcal{I},\mathcal{I}^{c}}$.
If the graph is a tree, BP can be used to produce the exact marginal
variances in a finite number of iterations. This largely corresponds
to applying the Cholesky factorization and Takahashi equations in
the case with no fill-in, which is also computationally cheap. In
the common case that the graph is not a tree, LBP must be used, which
is known to not converge for all models. \citet{Malioutov2006} show
that a sufficient condition for the convergence of the means and variances
is that the model is \textit{walk-summable,} a property including
for example models that have a diagonally dominant $\mathbf{Q}$.
However, only the means, and not the variances, are guaranteed to
converge to the true values, which together with the fact that many
models are not walk-summable limits the use of these methods. 

As a remedy, \citet{Liu2012} introduce a modified version named feedback
message passing (FMP), that first removes a number of ``feedback
nodes'' from the graph so that the remaining graph is cycle-free.
BP is then used to get the exact solution for this graph, that can
be passed back to the feedback nodes which in turn can now also get
the correct variances computed. In a final step, information from
the feedback nodes is used to compute the exact variances in the cycle-free
part of the graph. The method is exact and bears resemblance with
some of the methods from numerical linear algebra presented above,
but also becomes computationally intractable as the problem, and in
particular the number of required feedback nodes, becomes large. In
addition, a separate, faster, approximate FMP method is developed,
in which a smaller number of feedback nodes is selected so that the
remaining graph is no longer cycle-free, but at least walk-summable
or almost walk-summable. Approximate FMP is not exact, but is empirically
shown to produce reasonable approximations of the variances on some
small examples.

A number of papers look at sampling-based approaches for estimating
the selected inverse. The idea in \citet{Bekas2007} origins from
the paper by \citet{Hutchinson1990} and suggests estimating the matrix
diagonal as 
\begin{equation}
\hat{\boldsymbol{\sigma}}^{\mathbf{2}}=\left[\sum_{j=1}^{N_{s}}\mathbf{v}^{\left(j\right)}\odot\boldsymbol{\Sigma}\mathbf{v}^{\left(j\right)}\right]\oslash\left[\sum_{j=1}^{N_{s}}\mathbf{v}^{\left(j\right)}\odot\mathbf{v}^{\left(j\right)}\right],\label{eq:BekasEstimator}
\end{equation}
where $\odot$ and $\oslash$ means component-wise multiplication
and division of vectors respectively, where $\mathbf{v}^{\left(j\right)}$
is an $N$-dimensional random vector, for example a vector where each
element independently has value $1$ or $-1$ with equal probability,
and where $N_{s}$ is the number of sampled vectors. The method requires
the computation of $\boldsymbol{\Sigma}\mathbf{v}^{\left(j\right)}$,
which in our case can be done by solving $\mathbf{Q}\mathbf{z}=\mathbf{v}^{\left(j\right)}$
for $\mathbf{z}$, using PCG methods. The estimator in Eq.~(\ref{eq:BekasEstimator})
is unbiased, and it is also exact if the rows ($i$ and $j$) of $\mathbf{V}_{s}=\left[\mathbf{v}^{\left(1\right)},\ldots,\mathbf{v}^{\left(N_{s}\right)}\right]$
are orthogonal for all $i$ and $j$ for which $\boldsymbol{\Sigma}_{i,j}\neq0$.
If $\boldsymbol{\Sigma}$ is dense, this condition implies that $N_{s}=N$
is required for exactness (for example by choosing $\mathbf{v}^{\left(j\right)}=\mathbf{e}_{j}$),
which is not very helpful. However, this condition still motivates
choosing the columns of $\mathbf{V}_{s}$ deterministically, such
that the rows are non-orthogonal only for $i$ and $j$ such that
$\boldsymbol{\Sigma}_{i,j}$ is small. By assuming that the off-diagonal
elements of $\boldsymbol{\Sigma}$ decays with distance, \citet{Bekas2007}
motivates selecting $\mathbf{V}_{s}$ as a Hadamard matrix to get
a good approximation. The same sort of argument can be used to motivate
selecting $\mathbf{v}_{j}$ as probing vectors \citep{Tang2012},
but this requires first coloring the graph corresponding to $\mathbf{Q}^{p}$
for some suitable integer $p$. \citet{Malioutov2008} also use coloring
to select the rows of $\mathbf{V}_{s}$ as orthogonal for nodes that
are close, but in addition they provide a multiscale wavelet basis
for $\mathbf{V}_{s}$, that works better for long-range correlation
models and multiscale models.

\citet{Papandreou2010} develop an algorithm for fast sampling from
$\mathsf{N}\left(\mathbf{0},\mathbf{Q}^{-1}\right)$ when $\mathbf{Q}$
can be written as $\mathbf{G}^{T}\mathbf{G}+\mathbf{H}^{T}\mathbf{H}$,
for some sparse matrices $\mathbf{G}$ and $\mathbf{H}$, which is
a situation that often appears naturally when $\mathbf{Q}$ is a posterior
precision matrix, see for example the models in Section~\ref{sec:Results}.
In these cases, a sample can be produced as $\mathbf{x}^{\left(j\right)}=\mathbf{Q}^{-1}\left(\mathbf{G}^{T}\mathbf{z}_{1}+\mathbf{H}^{T}\mathbf{z}_{2}\right)$,
where $\mathbf{z}_{1}$ and $\mathbf{z}_{2}$ are standard normal
i.i.d. sampled vectors of appropriate lengths. For efficiency, the
PCG method can also here be used to solve the equation system with
respect to\textbf{ $\mathbf{Q}$}. A similar algorithm is provided
by \citet{Bhattacharya2016} to sample from the conditional posterior
in high-dimensional regression with Gaussian scale mixture priors.
Given $N_{s}$ independent samples of $\mathbf{x}$, denoted $\mathbf{X}=\left[\mathbf{x}^{\left(1\right)},\ldots,\mathbf{x}^{\left(N_{s}\right)}\right]$,
simple Monte Carlo (MC) estimators of $\boldsymbol{\Sigma}$ and $\sigma_{i}^{2}$
are
\begin{equation}
\hat{\boldsymbol{\Sigma}}_{MC}=\frac{1}{N_{s}}\sum_{j=1}^{N_{s}}\mathbf{x}^{\left(j\right)}\mathbf{x}^{\left(j\right)T}=\frac{1}{N_{s}}\mathbf{X}\mathbf{X}^{T},\,\,\,\,\,\,\,\text{\ensuremath{\sigma_{MC,i}^{2}=\frac{1}{N_{s}}\sum_{j=1}^{N_{s}}\left(x_{i}^{\left(j\right)}\right)^{2}},}\label{eq:MCestimator}
\end{equation}
which are further explored in \citet{Papandreou2011}. The estimators
follow scaled Wishart and chi-squared distributions with $N_{s}$
degrees of freedom, $\hat{\boldsymbol{\Sigma}}_{MC}\sim\frac{1}{N_{s}}\mathsf{Wishart}\left(\boldsymbol{\Sigma},N_{s}\right)$
and $\hat{\sigma}_{MC,i}^{2}\sim\frac{\sigma_{i}^{2}}{N_{s}}\chi_{N_{s}}^{2}$,
and are thus unbiased, see for example \citet[Chapter 3]{Mardia1979}.
By defining the relative error with respect to the true marginal variances
of the second estimator as $r_{MC,i}=\nicefrac{\left(\hat{\sigma}_{MC,i}^{2}-\sigma_{i}^{2}\right)}{\sigma_{i}^{2}}$
and the relative root-mean-square error (relative RMSE) as $RMSE_{MC,i}=\sqrt{E\left[r_{MC,i}^{2}\right]}$,
the unbiasedness and the variance of a $\chi^{2}$-distributed variable
gives $RMSE_{MC,i}=\sqrt{Var\left[\nicefrac{\hat{\sigma}_{MC,i}^{2}}{\sigma_{i}^{2}}\right]}=\sqrt{2/N_{s}}$.
This means for example $20\%$ relative RMSE when using $N_{s}=50$
samples. Note that the relative RMSE does not depend on the true variances.
The MC estimator provides a simple way to estimate $\boldsymbol{\Sigma}_{\mathcal{S}}$
for any reasonably sized index set $\mathcal{S}$. The computational
bottleneck is usually in producing the samples $\mathbf{X}$, and
given these, the additional computational costs are very low in both
time and memory. At the same time, the estimator is a bit simplistic
in the sense that information about the distribution encoded in the
precision matrix $\mathbf{Q}$ is discarded when only using the samples
$\mathbf{X}$. We therefore propose an improved Rao-Blackwellized
MC estimator in the next section. 

\section{Methods\label{sec:Methods}}

\subsection{Rao-Blackwellized Monte Carlo}

~\\
The simple, yet effective, idea of this section will be to improve
the MC estimator \citep{Papandreou2010,Papandreou2011}, by using
the fact that the precision matrix is known. We will start by deriving
what we call the simple Rao-Blackwellized Monte Carlo (simple RBMC)
estimator and then propose a number of improvements resulting in what
we will refer to as the block RBMC estimator. 

We derive the simple RBMC approximation for $\sigma_{i}^{2}$ by using
the law of total variance
\begin{eqnarray}
Var\left(x_{i}\right) & = & E\left[Var\left(x_{i}|\mathbf{x}_{-i}\right)\right]+Var\left[E\left(x_{i}|\mathbf{x}_{-i}\right)\right]=Q_{i,i}^{-1}+Var\left[-Q_{i,i}^{-1}\mathbf{Q}_{i,-i}\mathbf{x}_{-i}\right]\nonumber \\
 & \approx & Q_{i,i}^{-1}+\frac{1}{N_{s}}\sum_{j=1}^{N_{s}}\left(Q_{i,i}^{-1}\mathbf{Q}_{i,-i}^{\,}\mathbf{x}_{-i}^{\left(j\right)}\right)^{2}=\hat{\sigma}_{i|-i}^{2},\label{eq:RBMCestimator}
\end{eqnarray}
with $-i$ denoting all indices but $i$ and the notation $\cdot|-i$
denotes that the part of the variance that comes from indices $-i$
are approximated using MC samples. This estimator also follows a (translated
and scaled) chi-squared distribution, $\hat{\sigma}_{i|-i}^{2}\sim Q_{i,i}^{-1}+\frac{\sigma_{i}^{2}-Q_{i,i}^{-1}}{N_{s}}\chi_{N_{s}}^{2}$,
and is clearly unbiased, see Section~\ref{subsec:Approximation-variance-and}.
The relative RMSE is $\left(1-Q_{i,i}^{-1}/\sigma_{i}^{2}\right)\sqrt{2/N_{s}}$,
so the reduction in relative RMSE by using this estimator instead
of the MC estimator is $Q_{i,i}^{-1}/\sigma_{i}^{2}$. The logic here
is that the closer $Q_{i,i}^{-1}$ is to $\sigma_{i}^{2}$ (they become
equal when $\mathbf{Q}$ is diagonal) the smaller the error becomes,
as a larger portion of the variance is then explained by $Q_{i,i}^{-1}$.
So for a GMRF which has close to independent elements the simple RBMC
approximation is much better, and as the dependence between the elements
increases the difference in relative RMSE between the methods decreases,
but RBMC is always better.

Let $\mathcal{D}\left(\mathbf{Q}\right)$ denote the diagonal matrix
with the same diagonal as $\mathbf{Q}$. As the expression $\mathbf{Q}_{i,-i}\mathbf{x}_{-i}^{\left(j\right)}$
can be compactly computed for all $i$ and $j$ as $\left(\mathbf{Q}-\mathcal{D}\left(\mathbf{Q}\right)\right)\mathbf{X}$,
it is clear that, given $\mathbf{X}$, the computational cost of the
simple RBMC estimator for all marginal variances is dominated by $N_{s}$
(sparse) matrix-vector-multiplications of size $N$. This is normally
cheap, more precisely $O\left(N\cdot N_{s}\right)$ when the number
of non-zero elements in each row of $\mathbf{Q}$ does not depend
on $N$.

The reduction in error compared to the MC estimator can be understood
from the two terms in Eq.~\ref{eq:RBMCestimator}, where the first
one is now computed exactly and only the second one is approximated
using MC samples. The estimator can be further improved by enlarging
the set of nodes for which covariances are computed exactly. A more
general RBMC estimator is written as
\begin{equation}
\hat{\boldsymbol{\Sigma}}_{\mathcal{S}|\mathcal{I}^{c}}=\left[Var\left(\mathbf{x}_{\mathcal{I}}|\mathbf{x}_{\mathcal{I}^{c}}\right)+\widehat{Var}\left[E\left(\mathbf{x}_{\mathcal{I}}|\mathbf{x}_{\mathcal{I}^{c}}\right)\right]\right]_{\mathcal{S}}=\left[\mathbf{Q}_{\mathcal{I},\mathcal{I}}^{-1}+\frac{1}{N_{s}}\sum_{j=1}^{N_{s}}\boldsymbol{\kappa}_{\mathcal{I}}^{\left(j\right)}\boldsymbol{\kappa}_{\mathcal{I}}^{\left(j\right)T}\right]_{\mathcal{S}},\label{eq:RBMCestimator-general}
\end{equation}
where $\boldsymbol{\kappa}_{\mathcal{I}}^{\left(j\right)}=\mathbf{Q}_{\mathcal{I},\mathcal{I}}^{-1}\mathbf{Q}_{\mathcal{I},\mathcal{I}^{c}}\mathbf{x}_{\mathcal{I}^{c}}^{\left(j\right)}$,
$\mathcal{I}$ is a subset of all nodes and the operator $\left[\cdot\right]_{\mathcal{S}}$
extracts the elements in $\mathcal{S}\subseteq\left\{ \left(i,j\right);i,j\in\mathcal{I}\right\} $.
Also this estimator can easily be shown to be unbiased, and follows
a Wishart distribution, see Section~\ref{subsec:Approximation-variance-and}.
The set of nodes $\mathcal{I}$ should be thought of as a spatial
enclosure of the nodes in $\mathcal{S}$, and assuming spatial dependence
that decays with distance, the approximation will be better the further
inside the interior of $\mathcal{I}$ the nodes in $\mathcal{S}$
are. If $\mathcal{I}$ is chosen as the whole domain, we get the exact
inverse. There is thus a tradeoff between computing cost and error
when selecting $\mathcal{I}$ for this estimator; a larger enclosure
size $M=\left|\mathcal{I}\right|$ leads to smaller error, but also
to heavier computations since Eq.~(\ref{eq:RBMCestimator-general})
contains an inverse of an $M\times M$ matrix. We illustrate the error
reduction with an example.

\begin{figure}[H]
\includegraphics[width=0.5\linewidth]{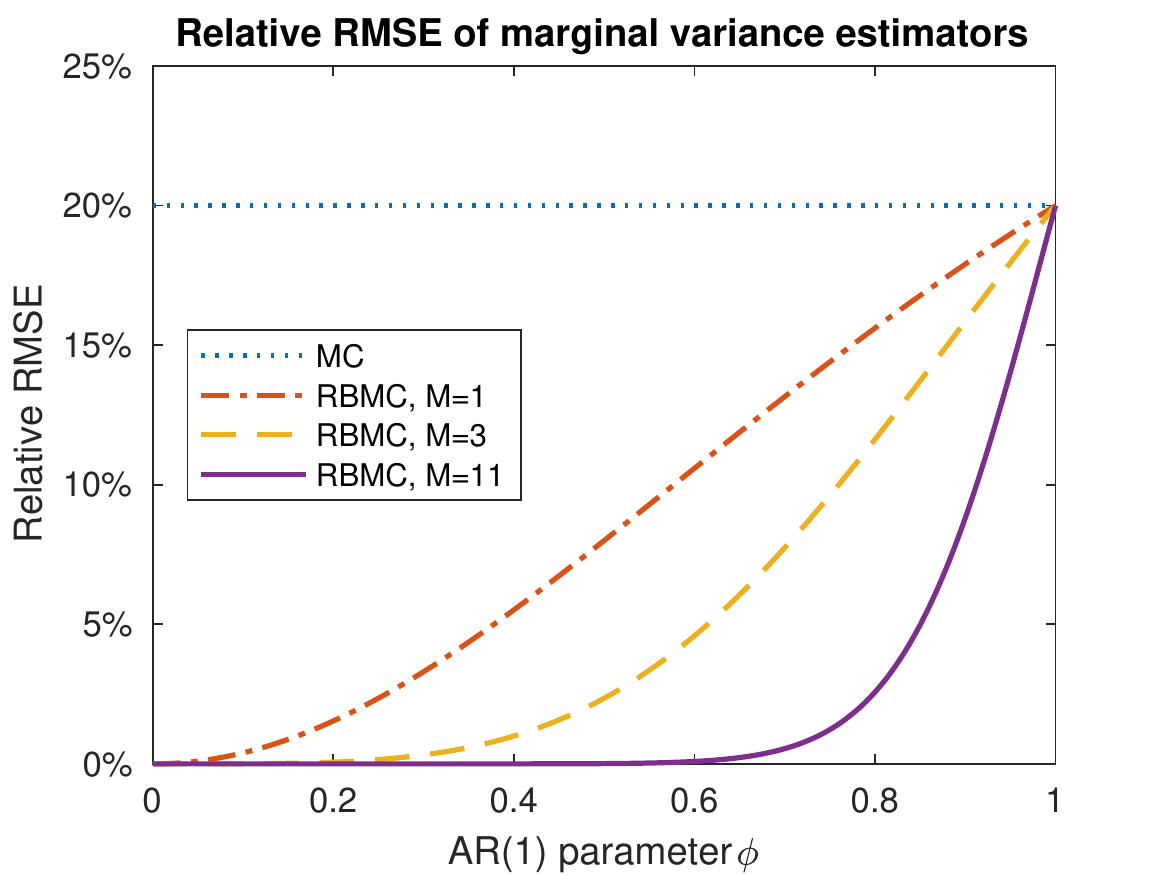}\caption{Relative RMSE of marginal variance estimators $\left(RMSE=\sqrt{E\left[\left(\nicefrac{\left(\hat{\sigma}_{i}^{2}-\sigma_{i}^{2}\right)}{\sigma_{i}^{2}}\right)^{2}\right]}\right)$
of the MC estimator $\left(RMSE=\sqrt{2/N_{s}}\right)$ and RBMC estimators
$\hat{\sigma}_{i|-i}^{2}$ $\left(M=1\right)$, $\hat{\sigma}_{i|-\left\{ i-1,i,i+1\right\} }^{2}$
$\left(M=3\right)$ and $\hat{\sigma}_{i|-\left\{ i-5,\ldots,i+5\right\} }^{2}$
$\left(M=11\right)$ $\left(RMSE=\frac{2\phi^{M+1}}{1+\phi^{M+1}}\sqrt{2/N_{s}}\right)$
for the AR$(1)$-model as a function of the AR-parameter $\phi$ and
$N_{s}=50$.\label{fig:AR1error}}
\end{figure}

A stationary AR$(1)$-process is possibly the simplest example of
a GMRF and can be defined as $x_{i}=\phi x_{i-1}+\varepsilon_{i},$
with $\varepsilon_{i}\sim\mathsf{N}\left(0,1\right)$ which gives
marginal variances $\left(1-\phi^{2}\right)^{-1}$. Ignoring the boundaries,
the precision matrix \textbf{$\mathbf{Q}$} for the $AR(1)$ is tridiagonal,
with $1+\phi^{2}$ on the diagonal and $-\phi$ on the super-/sub-diagonal,
while $\boldsymbol{\Sigma}$ is full. Fig.~\ref{fig:AR1error} depicts
the analytically derived relative RMSE for the MC estimator and three
different RBMC estimators for this model, plotted as a function of
$\phi$ when $N_{s}=50$, illustrating how the RBMC error increases
when the spatial dependence is increased and decreases when $M$ is
increased.

\begin{figure}[H]
\includegraphics[width=1\linewidth]{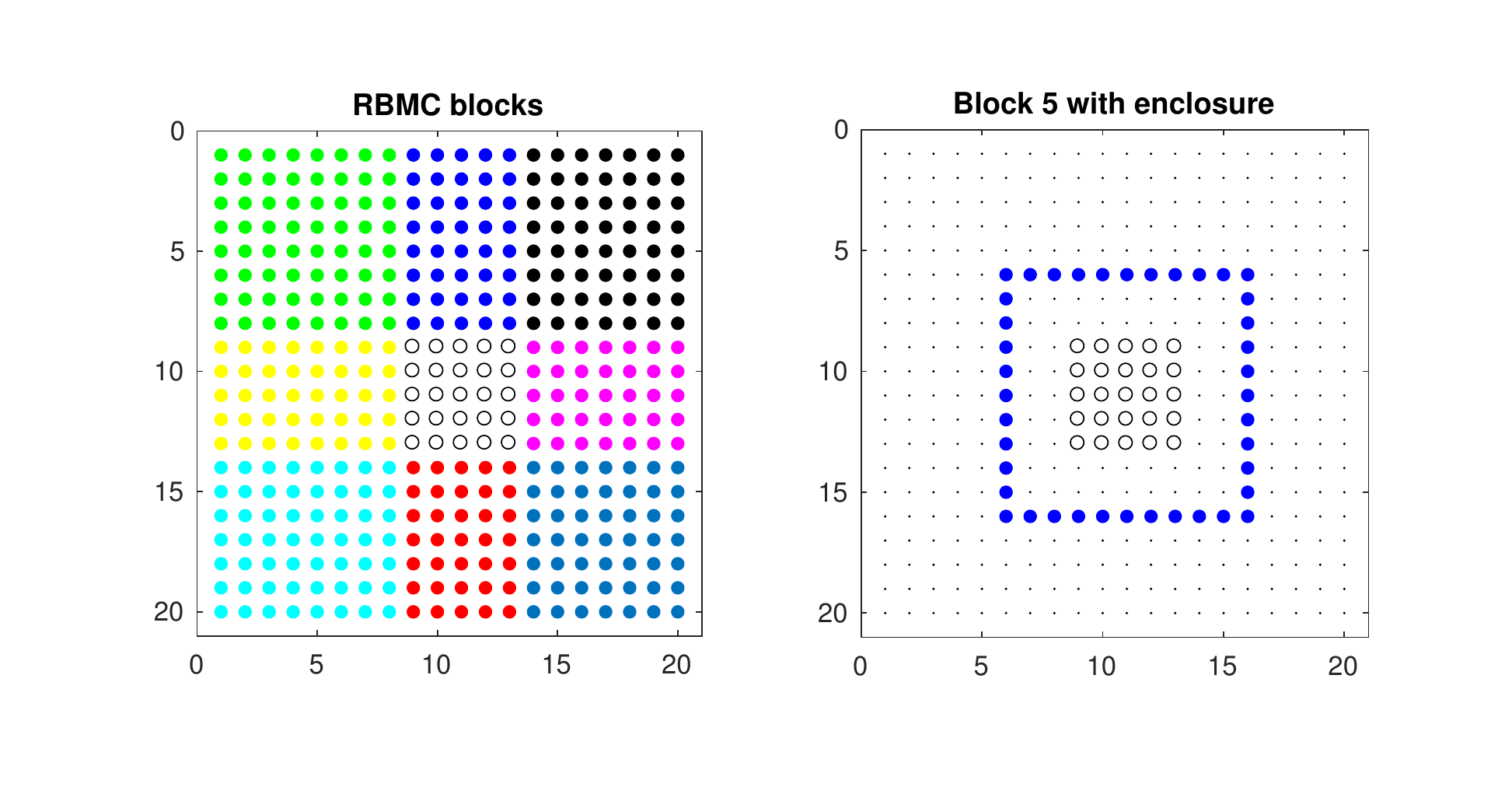}\caption{Disjoint blocks $\left\{ \mathcal{Y}_{1},\ldots,\mathcal{Y}_{9}\right\} $
for block RBMC in different colors (left) and block 5 together with
its spatial enclosure $\mathcal{I}\left(\mathcal{Y}_{5}\right)$ ,
that is, all nodes inside of the filled square (right).\label{fig:blockDisp}}
\end{figure}

\subsubsection{Block RBMC}

~\\
To compute the matrix diagonal using the improved RBMC estimator,
we could use the following strategy. For each element $i$ we select
a spatial enclosure $\mathcal{I}(i)$ of size $M$ and compute $\hat{\sigma}_{i|\mathcal{I}\left(i\right)^{c}}^{2}$
using Eq.~(\ref{eq:RBMCestimator-general}). The computational bottleneck
of the method will then be in the $N$ Cholesky factorizations of
$M\times M$-matrices, needed to compute $\mathbf{Q}_{\mathcal{I}\left(i\right),\mathcal{I}\left(i\right)}^{-1}$
and $\boldsymbol{\kappa}_{\mathcal{I}\left(i\right)}^{\left(j\right)}$
for each $i$. In practice, this strategy might lead to substantial
overhead costs when $N$ is large. In the numerical experiments in
Section~\ref{sec:Results}, we will therefore resort a to different
strategy which we refer to as block RBMC. In block RBMC, we partition
the domain into $N_{b}$ disjoint sets $\left\{ \mathcal{Y}_{1},\ldots,\mathcal{Y}_{N_{b}}\right\} $
and compute $\hat{\boldsymbol{\Sigma}}_{\mathcal{Y}_{_{i}}|\mathcal{I}\left(\mathcal{Y}_{i}\right)^{c}}$
for each block $i$ using Eq.~(\ref{eq:RBMCestimator-general}).
An example for a 20 by 20 lattice and 9 blocks is displayed in Fig.~\ref{fig:blockDisp}.

When we are only interested in the covariance matrix diagonal, the
important elements of Eq.~(\ref{eq:RBMCestimator-general}) can be
computed more efficiently. We start by reordering the nodes in the
spatial enclosure $\mathcal{I}\left(\mathcal{Y}_{i}\right)$ using
constrained approximate minimum degree (CAMD) reordering \citep{Liu1989,Amestoy1996},
such that the block nodes $\mathcal{Y}_{i}$ comes last. Assuming
$\left|\mathcal{I}\left(\mathcal{Y}_{i}\right)\right|$ is reasonably
small, we can then compute the Cholesky factor of $\mathbf{Q}_{\mathcal{I}\left(\mathcal{Y}_{i}\right),\mathcal{I}\left(\mathcal{Y}_{i}\right)}$
cheaply and also the sparse inverse of $\mathbf{Q}_{\mathcal{I}\left(\mathcal{Y}_{i}\right),\mathcal{I}\left(\mathcal{Y}_{i}\right)}$
using the Takahashi equations (see Eq.~(\ref{eq:Takahashi})). Since
we placed the block nodes last we do not need to iterate backwards
the whole way to $i=1$ but can break as soon as the variances of
the block nodes are computed. We thus have computed the first term
of Eq.~(\ref{eq:RBMCestimator-general}) and to obtain the second
term we can compute $\boldsymbol{\kappa}_{\mathcal{I}\left(\mathcal{Y}_{i}\right)}^{\left(j\right)}$
for all $j$ with forward and backward substitution using the Cholesky
factor. The block RBMC method is summarized in Algorithm \ref{Alg:Block RBMC}.
As presented there, block RBMC can be used to compute for example
the covariance matrix diagonal. For some other possible choices of
$\mathcal{S}$, trivial extensions to the algorithm could be required,
including making the blocks overlapping and computing additional elements
in the Takahashi equation step.

\begin{algorithm}[H]
\begin{algorithmic}[1]
\Require Precision matrix $\mathbf{Q}$, $N_s$ Gaussian samples $\mathbf{X}$, \newline blocks $\left\{ \mathcal{Y}_{1},\ldots,\mathcal{Y}_{N_{b}}\right\}$, and block enclosures $\left\{ \mathcal{I}\left(\mathcal{Y}_{1}\right),\ldots,\mathcal{I}\left(\mathcal{Y}_{N_{b}}\right)\right\}$
\vspace{1mm}
\For{ $i=1$ to $N_b$}
\vspace{1mm}        
\State Reorder the nodes in $\mathcal{I}\left(\mathcal{Y}_{i}\right)$ using CAMD such that $\mathcal{Y}_{i}$  comes last
\vspace{1mm}
\State Compute $\mathbf{L}_{\mathcal{I}\left(\mathcal{Y}_{i}\right)}$ as the Cholesky factor of $\mathbf{Q}_{\mathcal{I}\left(\mathcal{Y}_{i}\right),\mathcal{I}\left(\mathcal{Y}_{i}\right)}$
\vspace{1mm}
\For{ $j=\left|\mathcal{I}\left(\mathcal{Y}_{i}\right)\right|$ to $\left|\mathcal{I}\left(\mathcal{Y}_{i}\right)\right|-\left|\mathcal{Y}_{i}\right|+1$}
\vspace{1mm}
\State Use the Takahashi equations to compute sparse elements $(j,k)$\newline \color{white}.\color{black}\hspace{1cm} of $\mathbf{Q}_{\mathcal{I}\left(\mathcal{Y}_{i}\right),\mathcal{I}\left(\mathcal{Y}_{i}\right)}^{-1}$, for $k\geq j$
\vspace{1mm}
\EndFor
\vspace{1mm}
\For{ $j=1$ to $N_s$}
\vspace{1mm}
\State Solve $\mathbf{L}_{\mathcal{I}\left(\mathcal{Y}_{i}\right)} \tilde{\boldsymbol{\kappa}}=\mathbf{Q}_{\mathcal{I}\left(\mathcal{Y}_{i}\right),\mathcal{I}\left(\mathcal{Y}_{i}\right)^{c}}\mathbf{x}_{\mathcal{I}\left(\mathcal{Y}_{i}\right)^{c}}^{\left(j\right)}$ \hspace{1mm}for $\tilde{\boldsymbol{\kappa}}$
\vspace{1mm}
\State Solve $\mathbf{L}_{\mathcal{I}\left(\mathcal{Y}_{i}\right)}^{T} \boldsymbol{\kappa}_{\mathcal{I}\left(\mathcal{Y}_{i}\right)}^{\left(j\right)}=\tilde{\boldsymbol{\kappa}}$\hspace{1mm} for $\boldsymbol{\kappa}_{\mathcal{I}\left(\mathcal{Y}_{i}\right)}^{\left(j\right)}$
\vspace{1mm}
\EndFor
\vspace{1mm}
\State Compute selected covariances in block $\mathcal{Y}_{i}$ using Eq.\ (\ref{eq:RBMCestimator-general})
\vspace{1mm}
\State Optionally compute the uncertainty measures using Eq.~(\ref{eq:RBMCestDistr})
\vspace{1mm}
\EndFor
\end{algorithmic}\caption{Block RBMC\label{Alg:Block RBMC}}
\end{algorithm}

In the results presented in Section~\ref{sec:Results}, the choices
of blocks $\mathcal{Y}_{i}$ and enclosures $\mathcal{I}\left(\mathcal{Y}_{i}\right)$
will for simplicity be done correspondingly to how the blocks are
chosen for the iterative interface method presented in Section~\ref{subsec:Iterative-interface-method}
below. That is, for each block we select $\mathcal{Y}_{i}$ as the
smallest rectangle (or cuboid in the 3D case) that contains all nodes
in $\mathcal{Z}_{i}$ and $\mathcal{I}\left(\mathcal{Y}_{i}\right)=\mathcal{I}\left(\mathcal{W}_{i}\right)$
(see definitions of $\mathcal{Z}_{i}$ and $\mathcal{I}\left(\mathcal{W}_{i}\right)$
in Section~\ref{subsec:Iterative-interface-method}, and the illustration
in Fig.~\ref{fig:interfaceDisp}). This seems to be a pragmatic choice
of blocks for the RBMC method in practice.

\subsubsection{Approximation variance and confidence bounds\label{subsec:Approximation-variance-and}}

~\\
Precise estimates of the variance and uncertainty bounds of the different
RBMC estimators can be cheaply obtained by noting that the estimator
in Eq.~(\ref{eq:RBMCestimator-general}) follows a Wishart distribution.
See \citet[Chapter 3]{Mardia1979} for some fundamental properties
that connects the Wishart, Gaussian, and $\chi^{2}$-distributions.
It can be seen that $\boldsymbol{\kappa}_{\mathcal{I}}^{\left(j\right)}$
in Eq.~(\ref{eq:RBMCestimator-general}) is multivariate normal with
mean zero and covariance matrix $\boldsymbol{\Sigma}_{\mathcal{I},\mathcal{I}}-\mathbf{Q}_{\mathcal{I},\mathcal{I}}^{-1}$
since $\mathbf{x}^{\left(j\right)}$ is normal and since the law of
total variance gives that $\boldsymbol{\Sigma}_{\mathcal{I},\mathcal{I}}=\mathbf{Q}_{\mathcal{I},\mathcal{I}}^{-1}+Var\left(-\boldsymbol{\kappa}_{\mathcal{I}}^{\left(j\right)}\right)$.
It thereby follows that 
\begin{equation}
\hat{\boldsymbol{\Sigma}}_{\left(\mathcal{I},\mathcal{I}\right)|\mathcal{I}^{c}}\sim\mathbf{Q}_{\mathcal{I},\mathcal{I}}^{-1}+\frac{1}{N_{s}}\mathsf{Wishart}\left(\boldsymbol{\Sigma}_{\mathcal{I},\mathcal{I}}-\mathbf{Q}_{\mathcal{I},\mathcal{I}}^{-1},N_{s}\right),\label{eq:RBMCestDistr}
\end{equation}
 and taking the mean directly shows the unbiasedness of the estimator. 

We thus know the analytical distribution of the different RBMC estimators.
This can be used to compute uncertainty measures such as the variances
and confidence bounds of the different elements, apart from that we
do not know $\boldsymbol{\Sigma}_{\mathcal{I},\mathcal{I}}$ which
is in fact what we are trying to estimate. However, if $\hat{\boldsymbol{\Sigma}}_{\left(\mathcal{I},\mathcal{I}\right)|\mathcal{I}^{c}}$
is a reasonably good estimate of $\boldsymbol{\Sigma}_{\mathcal{I},\mathcal{I}}$,
then plugging it into Eq.~(\ref{eq:RBMCestDistr}) instead of $\boldsymbol{\Sigma}_{\mathcal{I},\mathcal{I}}$
gives good estimates also of the uncertainty measures, as shown empirically
in Section~\ref{sec:Results}. Note that $\hat{\boldsymbol{\Sigma}}_{\left(\mathcal{I},\mathcal{I}\right)|\mathcal{I}^{c}}-\mathbf{Q}_{\mathcal{I},\mathcal{I}}^{-1}$
is positive definite, due to the construction in Eq~(\ref{eq:RBMCestimator-general}).
Also note that both $\hat{\boldsymbol{\Sigma}}_{\left(\mathcal{I},\mathcal{I}\right)|\mathcal{I}^{c}}$
and $\mathbf{Q}_{\mathcal{I},\mathcal{I}}^{-1}$ are already computed
for selected elements in Algorithm~\ref{Alg:Block RBMC}, so computing
the uncertainty measures generates very little additional computational
cost.

As an example, we give explicit uncertainty measures for the RBMC
estimates of the elements of covariance matrix diagonal $\hat{\sigma}_{i|\mathcal{I}^{c}}^{2}$.
These can be derived by noting that the diagonal elements of a Wishart
distributed matrix are $\chi^{2}$-distributed, so 
\begin{equation}
\hat{\sigma}_{i|\mathcal{I}^{c}}^{2}\sim\left[\mathbf{Q}_{\mathcal{I},\mathcal{I}}^{-1}\right]_{i,i}+\frac{1}{N_{s}}\left(\sigma_{i}^{2}-\left[\mathbf{Q}_{\mathcal{I},\mathcal{I}}^{-1}\right]_{i,i}\right)\chi_{N_{s}}^{2},\,\,\,\,\,\,\,Var\left(\hat{\sigma}_{i|\mathcal{I}^{c}}^{2}\right)=\frac{2}{N_{s}}\left(\sigma_{i}^{2}-\left[\mathbf{Q}_{\mathcal{I},\mathcal{I}}^{-1}\right]_{i,i}\right)^{2},\label{eq:RBMCestDiagDistr}
\end{equation}
and the quantiles of the $\chi^{2}$-distribution can directly be
used to compute confidence intervals (CIs). In practice the uncertainty
measures can be approximated using $\sigma_{i}^{2}=\hat{\sigma}_{i|\mathcal{I}^{c}}^{2}$.

\subsection{Iterative interface method\label{subsec:Iterative-interface-method}}

\begin{figure}[H]
\includegraphics[width=1\linewidth]{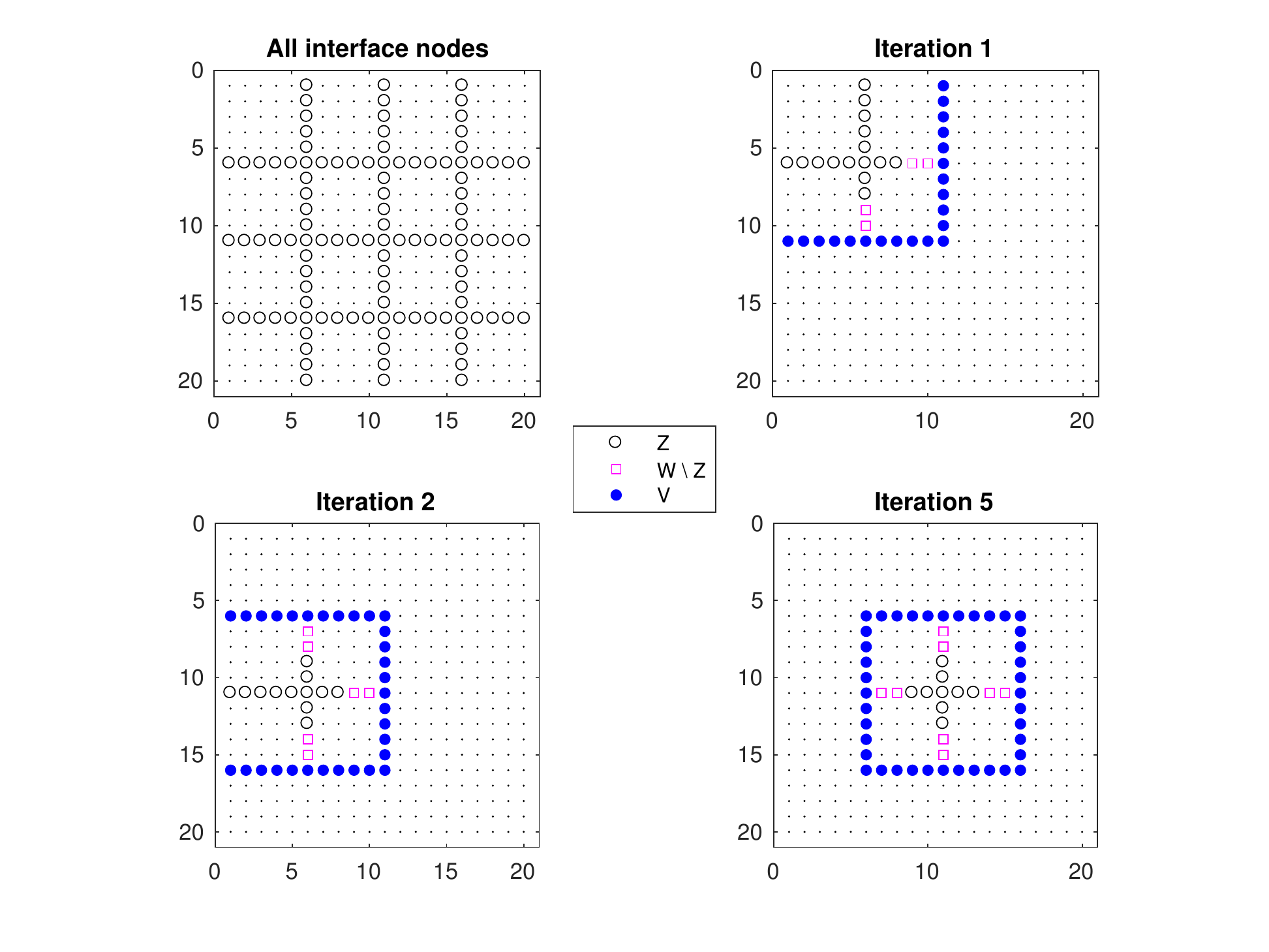}\caption{The iterative interface method illustrated for the $20$ by $20$
lattice with 9 subblocks. All interface nodes (top left) and those
updated in the iterations 1, 2 and 5, divided into the sets $\mathcal{W}$,
$\mathcal{V}$ and $\mathcal{Z}$. In addition, $\mathcal{I}\left(\mathcal{W}\right)$
denotes all nodes inside the frame and $\mathcal{U}$ denotes all
nodes on and inside the frame $\mathcal{V}$. \label{fig:interfaceDisp}}
\end{figure}

~\\
In this section we introduce a method that can be used to further
improve the RBMC covariance estimates by iterating over certain subdomains,
which we call \emph{interfaces}. For the ease of presentation, we
will here assume that we have a GMRF defined on a 2D lattice with
nearest neighbor Markov structure (5-point-stencil), but it is straightforward
to extend this to 3D (we provide numerical results for this case in
Section~\ref{sec:Results}) and also possible for other types of
domains or Markov structures. The underlying idea can be explained
using Fig.~\ref{fig:interfaceDisp}, which depicts a $20$ by $20$
lattice with all the interface nodes marked with unfilled dots in
the top left graph. The other three graphs illustrate situations in
which a subset of interface nodes $\mathcal{W}_{i}$ (unfilled) have
been enclosed within a frame of other interface nodes $\mathcal{V}_{i}$
(filled). The nodes in $\mathcal{W}_{i}$ are divided into an inner
set $\mathcal{Z}_{i}$ (unfilled circles) and an outer set $\mathcal{W}_{i}\setminus\mathcal{Z}_{i}$
(unfilled squares) for reasons that will be apparent shortly. We also
use the notation $\mathcal{I}\left(\mathcal{W}_{i}\right)$ for all
nodes within the frame and $\mathcal{U}_{i}=\mathcal{I}\left(\mathcal{W}_{i}\right)\cup\mathcal{V}_{i}$,
that is, $\mathcal{U}_{i}$ are all nodes on and inside the frame.
Because of the Markov assumption, if we would know the covariance
matrix $\boldsymbol{\Sigma}_{\mathcal{V}_{i},\mathcal{V}_{i}}$ of
the frame, we could compute the covariance matrix of the inner nodes
$\boldsymbol{\Sigma}_{\mathcal{W}_{i},\mathcal{W}_{i}}$ without having
to consider the distribution outside of the frame. The basic idea
is therefore to iterate between interface subdomains, as the three
illustrated, and in each step compute the covariances of the inner
nodes $\text{\ensuremath{\mathcal{W}} }_{i}$ based on the covariances
of the frame $\mathcal{V}_{i}$ and $\mathbf{Q}$ (for the example
in Fig.~\ref{fig:interfaceDisp}, nine steps are required to iterate
through all interface nodes once).

The algorithm, summarized in Algorithm~\ref{Alg:Iterative interface method},
can be divided into three phases. In the first phase, starting values
are computed using a slightly modified version of the block RBMC method,
in which the full covariance matrix of the innermost nodes $\mathcal{Z}_{i}$
are computed together with the cross-covariances between $\mathcal{Z}_{i}$
and $\mathcal{W}_{i}\setminus\mathcal{Z}_{i}$. The starting variances
of the nodes in $\mathcal{W}_{i}\setminus\mathcal{Z}_{i}$ are however
estimated in a different block where these nodes are further inside
the frame, which leads to smaller error (consider for example the
two bottommost square nodes in iteration 1 of Fig.~\ref{fig:interfaceDisp},
which are more centrally located in iteration 2). In the second phase
the algorithm iterates $N_{iter}$ times over all $N_{b}$ blocks
and computes $\boldsymbol{\Sigma}_{\mathcal{W}_{i},\mathcal{W}_{i}}$
each time treating $\boldsymbol{\Sigma}_{\mathcal{V}_{i},\mathcal{V}_{i}}$
as known. In the final phase all selected covariances (not only those
that happen to belong to interface nodes) are computed using the Takahashi
equations with the modification that the covariances on the frame,
$\boldsymbol{\Sigma}_{\mathcal{V}_{i},\mathcal{V}_{i}}$, are treated
as known. More formal derivations and motivations of the different
steps in Algorithm~\ref{Alg:Iterative interface method} are given
in the following subsection.

\subsubsection{Algorithm derivation}

~\\
Note that every step in Algorithm~\ref{Alg:Iterative interface method}
is done within the context of a single subblock/interface, so here
we drop the subindex $i$ from all sets, for readability. Consider
the case when we are interested in computing the dense covariance
matrix $\boldsymbol{\Sigma}_{\mathcal{W},\mathcal{W}}$ knowing $\boldsymbol{\Sigma}_{\mathcal{V},\mathcal{V}}$
and using that $p\left(\mathbf{x}_{\mathcal{W}}|\mathbf{x}_{\mathcal{V}}\,,\mathbf{x}_{\mathcal{U}^{c}}\right)=p\left(\mathbf{x}_{\mathcal{W}}|\mathbf{x}_{\mathcal{V}}\right).$
For computational efficiency, first reorder the nodes in $\mathbf{Q}_{\mathcal{I}\left(\mathcal{W}\right),\mathcal{I}\left(\mathcal{W}\right)}$
such that $\mathcal{W}$ comes last, using CAMD. Now, similar to when
the RBMC method was derived
\begin{align}
Var\left(\mathbf{x}_{\mathcal{I}\left(\mathcal{W}\right)}\right)=\boldsymbol{\Sigma}_{\mathcal{I}\left(\mathcal{W}\right),\mathcal{I}\left(\mathcal{W}\right)} & =E\left[Var\left(\mathbf{x}_{\mathcal{I}\left(\mathcal{W}\right)}|\mathbf{x}_{\mathcal{V}}\right)\right]+Var\left[E\left(\mathbf{x}_{\mathcal{I}\left(\mathcal{W}\right)}|\mathbf{x}_{\mathcal{V}}\right)\right]\nonumber \\
 & =\mathbf{Q}_{\mathcal{I}\left(\mathcal{W}\right),\mathcal{I}\left(\mathcal{W}\right)}^{-1}+Var\left(\mathbf{Q}_{\mathcal{I}\left(\mathcal{W}\right),\mathcal{I}\left(\mathcal{W}\right)}^{-1}\mathbf{Q}_{\mathcal{I}\left(\mathcal{W}\right),\mathcal{V}}\mathbf{x}_{\mathcal{V}}\right)\nonumber \\
 & =\mathbf{L}^{-T}\mathbf{L}^{-1}+Var\left(\mathbf{L}^{-T}\mathbf{L}^{-1}\mathbf{Q}_{\mathcal{I}\left(\mathcal{W}\right),\mathcal{V}}\mathbf{x}_{\mathcal{V}}\right)\nonumber \\
 & =\mathbf{L}^{-T}\left(\mathbf{I}_{\left|\mathcal{I}\left(\mathcal{W}\right)\right|}+Var\left(\mathbf{L}^{-1}\mathbf{Q}_{\mathcal{I}\left(\mathcal{W}\right),\mathcal{V}}\mathbf{x}_{\mathcal{V}}\right)\right)\mathbf{L}^{-1}\nonumber \\
 & =\mathbf{L}^{-T}\left(\mathbf{I}_{\left|\mathcal{I}\left(\mathcal{W}\right)\right|}+\mathbf{M}\boldsymbol{\Sigma}_{V,V}\mathbf{M}^{T}\right)\mathbf{L}^{-1}\label{eq:Schur}
\end{align}
where $\mathbf{Q}_{\mathcal{I}\left(\mathcal{W}\right),\mathcal{I}\left(\mathcal{W}\right)}=\mathbf{LL}^{T}$
is the Cholesky decomposition and \textbf{$\mathbf{M}=\mathbf{L}^{-1}\mathbf{Q}_{\mathcal{I}\left(\mathcal{W}\right),\mathcal{V}}$}.
This equation provides a way to compute $\boldsymbol{\Sigma}_{\mathcal{I}\left(\mathcal{W}\right),\mathcal{I}\left(\mathcal{W}\right)}$
when $\boldsymbol{\Sigma}_{\mathcal{V},\mathcal{V}}$ is known, but
since we are only interested in the covariance matrix $\boldsymbol{\Sigma}_{\mathcal{W},\mathcal{W}}$,
this would be unnecessary. We divide the Cholesky factor $\mathbf{L}$
using the subsets $\mathcal{\widetilde{\mathcal{W}}}:=\mathcal{I}\left(\mathcal{W}\right)\setminus\mathcal{W}$
and $\mathcal{W}$ so that 
\begin{equation}
\mathbf{L}=\left[\begin{array}{cc}
\mathbf{L}_{\mathcal{\widetilde{\mathcal{W}}}}\\
\mathbf{L}_{\mathcal{W},\mathcal{\widetilde{\mathcal{W}}}} & \mathbf{L}_{\mathcal{W}}
\end{array}\right]\Rightarrow\mathbf{L}^{-1}=\left[\begin{array}{cc}
\mathbf{L}_{\mathcal{\widetilde{\mathcal{W}}}}^{-1}\\
-\mathbf{L}_{\mathcal{W}}^{-1}\mathbf{L}_{\mathcal{W},\mathcal{\widetilde{\mathcal{W}}}}\mathbf{L}_{\mathcal{\widetilde{\mathcal{W}}}}^{-1} & \mathbf{L}_{\mathcal{W}}^{-1}
\end{array}\right]=\left[\begin{array}{cc}
\mathbf{L}_{\mathcal{\widetilde{\mathcal{W}}}}^{-1}\\
-\mathbf{S} & \mathbf{L}_{\mathcal{W}}^{-1}
\end{array}\right],\label{eq:blockChol}
\end{equation}
and divide $\mathbf{M}$ as $\ \mathbf{M}^{T}=\left[\text{\textbf{M}}_{\mathcal{\widetilde{\mathcal{W}}}}^{T}\,\,\mathbf{M}_{\mathcal{W}}^{T}\right]$.
Eq.~(\ref{eq:Schur}) can now be written as
\begin{align}
\left[\begin{array}{cc}
\boldsymbol{\Sigma}_{\mathcal{\widetilde{\mathcal{W}}},\mathcal{\widetilde{\mathcal{W}}}} & \boldsymbol{\Sigma}_{\mathcal{\widetilde{\mathcal{W}}},\mathcal{W}}\\
\boldsymbol{\Sigma}_{\mathcal{W},\mathcal{\widetilde{\mathcal{W}}}} & \boldsymbol{\Sigma}_{\mathcal{W},\mathcal{W}}
\end{array}\right] & =\left[\begin{array}{cc}
\mathbf{L}_{\mathcal{\widetilde{\mathcal{W}}}}^{-T} & -\mathbf{S}\\
 & \mathbf{L}_{\mathcal{W}}^{-T}
\end{array}\right]\cdot\label{eq:Schur2}\\
 & \left(\left[\begin{array}{cc}
\mathbf{I}_{\left|\mathcal{\widetilde{\mathcal{W}}}\right|}+\mathbf{M}_{\mathcal{\widetilde{\mathcal{W}}}}\boldsymbol{\Sigma}_{\mathcal{V},\mathcal{V}}\mathbf{M}_{\mathcal{\widetilde{\mathcal{W}}}}^{T} & \mathbf{M}_{\mathcal{\widetilde{\mathcal{W}}}}\boldsymbol{\Sigma}_{\mathcal{V},\mathcal{V}}\mathbf{M}_{\mathcal{W}}^{T}\\
\mathbf{M}_{\mathcal{W}}\boldsymbol{\Sigma}_{\mathcal{V},\mathcal{V}}\mathbf{M}_{\mathcal{\widetilde{\mathcal{W}}}}^{T} & \mathbf{I}_{\left|\mathcal{W}\right|}+\mathbf{M}_{\mathcal{W}}\boldsymbol{\Sigma}_{\mathcal{V},\mathcal{V}}\mathbf{M}_{\mathcal{W}}^{T}
\end{array}\right]\right)\left[\begin{array}{cc}
\mathbf{L}_{\mathcal{\widetilde{\mathcal{W}}}}^{-1}\\
-\mathbf{S} & \mathbf{L}_{\mathcal{W}}^{-1}
\end{array}\right],\nonumber 
\end{align}
from which the bottom right block can be extracted. This gives
\begin{equation}
\boldsymbol{\Sigma}_{\mathcal{W},\mathcal{W}}=\mathbf{L}_{\mathcal{W}}^{-T}\left(\mathbf{I}_{\left|\mathcal{W}\right|}+\mathbf{M}_{\mathcal{W}}\boldsymbol{\Sigma}_{\mathcal{V},\mathcal{V}}\mathbf{M}_{\mathcal{W}}^{T}\right)\mathbf{L}_{\mathcal{W}}^{-1},\label{eq:updateW}
\end{equation}
and we now have a formula to update $\boldsymbol{\Sigma}_{W,W}$ given
$\boldsymbol{\Sigma}_{\mathcal{V},\mathcal{V}}$, which is used on
line 10 in Algorithm~\ref{Alg:Iterative interface method}. If $\boldsymbol{\Sigma}_{\mathcal{V},\mathcal{V}}$
is approximated by samples in Eq.~(\ref{eq:updateW}) we get the
RBMC estimator for the starting values in line 5
\begin{equation}
\boldsymbol{\hat{\Sigma}}_{\mathcal{W},\mathcal{W}}^{\text{start}}=\mathbf{L}_{\mathcal{W}}^{-T}\left(\mathbf{I}_{\left|\mathcal{W}\right|}+\frac{1}{N_{s}}\sum_{j=1}^{N_{s}}\left(\mathbf{M}_{\mathcal{W}}\mathbf{x}_{\mathcal{V}}^{\left(j\right)}\right)\left(\mathbf{M}_{\mathcal{W}}\mathbf{x}_{\mathcal{V}}^{\left(j\right)}\right)^{T}\right)\mathbf{L}_{\mathcal{W}}^{-1}.\label{eq:RBMCforW}
\end{equation}

\begin{algorithm}[H]
\begin{algorithmic}[1]
\Require Precision matrix $\mathbf{Q}$, Gaussian samples $\mathbf{X}$, \newline and node sets $\mathcal{W}_{i}$, $\mathcal{I}\left(\mathcal{W}_{i}\right)$, $\mathcal{V}_{i}$, $\mathcal{Z}_{i}$, $\mathcal{U}_{i}$ for all $i$
\vspace{1mm}
\For{ $i=1$ to $N_b$} \Comment{start phase one}
\vspace{1mm}        
\State Reorder the nodes in $\mathcal{I}\left(\mathcal{W}_{i}\right)$ using CAMD such that $\mathcal{W}_{i}$  comes last
\vspace{1mm}
\State Compute $\mathbf{L}$ as the Cholesky factor of $\mathbf{Q}_{\mathcal{I}\left(\mathcal{W}_{i}\right),\mathcal{I}\left(\mathcal{W}_{i}\right)}$ and extract $\mathbf{L}_{\mathcal{W}_{i}}$ as \newline \color{white}.\color{black}\hspace{4mm} the bottom right $\left|\mathcal{W}_{i}\right|\times\left|\mathcal{W}_{i}\right|$ block of $\mathbf{L}$
\vspace{1mm}
\State Compute $\mathbf{M}=\mathbf{L}^{-1}\mathbf{Q}_{\mathcal{I}\left(\mathcal{W}_{i}\right),\mathcal{V}_{i}}$ and extract $\mathbf{M}_{\mathcal{W}_{i}}$ as the last $\left|\mathcal{W}_{i}\right|$ rows of $\mathbf{M}$
\vspace{1mm}
\State Compute starting values as \newline \color{white}.\color{black}\hspace{4mm} $\boldsymbol{\hat{\Sigma}}_{\mathcal{W}_{i},\mathcal{W}_{i}}^{\text{start}}=\mathbf{L}_{\mathcal{W}_{i}}^{-T}\left(\mathbf{I}_{\left|\mathcal{W}_{i}\right|}+\frac{1}{N_{s}}\sum_{j=1}^{N_{s}}\left(\mathbf{M}_{\mathcal{W}_{i}}\mathbf{x}_{\mathcal{V}_{i}}^{\left(j\right)}\right)\left(\mathbf{M}_{\mathcal{W}_{i}}\mathbf{x}_{\mathcal{V}_{i}}^{\left(j\right)}\right)^{T}\right)\mathbf{L}_{\mathcal{W}_{i}}^{-1}$
\vspace{1mm}
\State Set $\boldsymbol{\hat{\Sigma}}_{\mathcal{S}}=\boldsymbol{\hat{\Sigma}}_{\mathcal{S}}^{\text{start}}$ for $\mathcal{S}=\left(\mathcal{Z}_{i}\times\mathcal{Z}_{i}\right)\cup\left(\left(\mathcal{W}_{i}\setminus\mathcal{Z}_{i}\right)\times\mathcal{Z}_{i}\right)\cup\left(\mathcal{Z}_{i}\times\left(\mathcal{W}_{i}\setminus\mathcal{Z}_{i}\right)\right)$
\vspace{1mm}
\EndFor
\vspace{1mm}
\For{ $j=1$ to $N_{iter}$} \Comment{start phase two}
\vspace{1mm}
\For{ $i=1$ to $N_b$}
\vspace{1mm}
\State Compute $\boldsymbol{\hat{\Sigma}}_{\mathcal{W}_{i},\mathcal{W}_{i}}=\mathbf{L}_{\mathcal{W}_{i}}^{-T}\left(\mathbf{I}_{\left|\mathcal{W}_{i}\right|}+\mathbf{M}_{\mathcal{W}_{i}}\boldsymbol{\hat{\Sigma}}_{\mathcal{V}_{i},\mathcal{V}_{i}}\mathbf{M}_{\mathcal{W}_{i}}^{T}\right)\mathbf{L}_{\mathcal{W}_{i}}^{-1}$
\vspace{1mm}
\EndFor
\vspace{1mm}
\EndFor
\vspace{1mm}
\For{ $i=1$ to $N_b$} \Comment{start phase three}
\vspace{1mm}        
\State Reorder the nodes in $\mathcal{U}_{i}$ using CAMD such that $\mathcal{V}_{i}$  comes last
\vspace{1mm}
\State Compute $\mathbf{L}_{\mathcal{U}_{i}}$ as the Cholesky factor of $\mathbf{Q}_{\mathcal{U}_{i},\mathcal{U}_{i}}$
\vspace{1mm}
\For{ $j=\left|\mathcal{I}\left(\mathcal{W}_{i}\right)\right|$ to $1$}
\vspace{1mm}
\State Use the Takahashi equations to compute sparse elements $(j,k)$\newline \color{white}.\color{black}\hspace{1cm} of $\boldsymbol{\hat{\Sigma}}_{\mathcal{U}_{i},\mathcal{U}_{i}}$, for $k\geq j$ treating the last block $\boldsymbol{\hat{\Sigma}}_{\mathcal{V}_{i},\mathcal{V}_{i}}$ as known.
\vspace{1mm}
\EndFor
\vspace{1mm}
\EndFor
\end{algorithmic}\caption{Iterative interface method\label{Alg:Iterative interface method}}
\end{algorithm}

\subsubsection{Convergence and error}

~\\
The hope is to bring the interface covariances closer to the exact
values in each iteration. However, the error will not converge to
zero in general since the necessary covariances between some nodes
in each frame can not be computed, and will instead be assumed to
be zero. For example, in the bottom right subgraph of Fig \ref{fig:interfaceDisp},
the covariance between the top left and bottom right nodes in the
frame will never be computed since these nodes are not in the same
$\mathcal{W}_{i}$ for any $i$. Still, for all interface nodes that
are close to each other the covariance will be computed in some block
$i$ and hopefully the approximation error from not computing the
covariance of distant nodes will be small. If not we can always increase
the sizes of the interface blocks, which increases the distance between
the nodes for which the covariance can not be computed. This will
however bring additional computational costs.

\subsection{Correcting for linear constraints}

~\\
A situation that occurs quite frequently in practice is that we have
some linear constraints $\mathbf{Ax}=\mathbf{e}$ on the GMRF \textbf{$\mathbf{x}$},
for example that $\sum_{i}x_{i}=0$, that is $\mathbf{A}=\mathbf{1}^{T}$
and $\mathbf{e}=0$ \citep{Isham2004}. In such a situation, \citet{Rue2007}
provide a general strategy to compute selected elements of the covariance
matrix $\boldsymbol{\Sigma}^{*}$ of $\mathbf{x}^{*}=\left(\mathbf{x}|\mathbf{Ax}=\mathbf{e}\right)$
using that 
\begin{equation}
\boldsymbol{\Sigma}^{*}=\boldsymbol{\Sigma}-\mathbf{Q}^{-1}\mathbf{A}^{T}\left(\mathbf{A}\mathbf{Q}^{-1}\mathbf{A}^{T}\right)^{-1}\mathbf{A}\mathbf{Q}^{-1}=\boldsymbol{\Sigma}-\mathbf{W}\left(\mathbf{A}\mathbf{W}\right)^{-1}\mathbf{W}^{T}=\boldsymbol{\Sigma}-\mathbf{C},\label{eq:correct-linear-constraints}
\end{equation}
where $\mathbf{W}=\mathbf{Q}^{-1}\mathbf{A}^{T}$ and $\mathbf{C}=\mathbf{W}\left(\mathbf{A}\mathbf{W}\right)^{-1}\mathbf{W}^{T}$.
If \textbf{$\mathbf{A}$} is of size $k\times N$, then the cost of
computing $\mathbf{W}$ is equal to that of solving $k$ equation
systems $\mathbf{Q}\mathbf{W}=\mathbf{A}^{T}$, which can be done
with PCG as explained earlier, as long as $k$ is reasonably small.
In this case, computing selected elements of $\mathbf{C}$ is also
cheap, requiring one $k\times k$ matrix inversion and an additional
$k\times k$ matrix-vector-multiplication per element. Thereby, given
an estimate $\hat{\boldsymbol{\Sigma}}_{\mathcal{S}}$ for some index
set $\mathcal{S}$ from any of the methods above, an estimate $\hat{\boldsymbol{\Sigma}}_{\mathcal{S}}^{*}=\hat{\boldsymbol{\Sigma}}_{\mathcal{S}}-\mathbf{C}_{\mathcal{S}}$
of selected elements of the covariance matrix of the constrained field
is straightforward to compute. As $\mathbf{C}_{\mathcal{S}}$ can
be computed exactly the variance of the estimator $\hat{\boldsymbol{\Sigma}}_{\mathcal{S}}^{*}$
is the same as the variance of $\hat{\boldsymbol{\Sigma}}_{\mathcal{S}}$.

When this method is used for the diagonal elements of $\boldsymbol{\Sigma}^{*}$,
the marginal variances, some attention should be drawn to the fact
that some estimates could become negative if $\hat{\boldsymbol{\Sigma}}_{i,i}<\mathbf{C}_{i,i}$
for some $i$. One possible approach to remedy this situation in practice
is to replace any negative estimates with the MC estimates computed
using samples from the constrained field itself (with subtracted mean),
which can be computed by correcting the samples from the original
field as $\mathbf{X}^{*}=\mathbf{X}-\mathbf{W}\left(\mathbf{A}\mathbf{W}\right)^{-1}\mathbf{A}\mathbf{X},$
see \citet[Algortihm 2.6]{Isham2004}.

\section{Results\label{sec:Results}}

In this section, we will investigate the performance of the introduced
methods for selected inversion empirically on various posterior precision
matrices. We first compare sampling-based methods on a simple theoretical
model and then consider a spatial model for neuroimaging and evaluate
all our methods using data from both simulated and real fMRI experiments.
All computations were performed on a Linux workstation with a 4-core
(8 threads) Intel Xeon E5-1620 processor at 3.5GHz and 128GB RAM.
The main part of the code was written in Matlab, but some (non-optimized)
C++ code was called for evaluating the Takahashi equations and the
SuiteSparse library \citep{SuiteSparse} was used for calling CAMD
and the functions for symbolic Cholesky factorization.

The first task consists in computing the covariance matrix diagonal
corresponding to the sparse posterior matrix of a simple model with
independent Gaussian measurements and a first order random walk prior
on the 3D lattice, that is $y_{i}\sim\mathsf{N}\left(x_{i},\lambda_{i}^{-1}\right)$
and $x_{i}-x_{j}\sim\mathsf{N}\left(0,1\right)$ for all adjacent
nodes $i$ and $j$ on the lattice a priori. $\lambda_{i}$ were uniformly
sampled on the interval $\left(0.1,0.2\right)$ for all $i$. The
posterior distribution for $\mathbf{x}|\mathbf{y}$ is a GMRF with
precision matrix $\mathbf{Q}=diag\left(\boldsymbol{\lambda}\right)+\mathbf{G}^{T}\mathbf{G}$,
with $\boldsymbol{\lambda}=\left[\lambda_{1},\ldots,\lambda_{N}\right]$
and $\mathbf{G}$ is a matrix with one row for every pair of adjacent
nodes $i$ and $j$, with $1$ in column $i$ and $-1$ in column
$j$. We compare the simple RBMC and block RBMC methods to the MC
and Hutchinson sampling-based methods in Table~\ref{tab:timingsSim}.
For each node $i$ we compute the relative error $r_{i}=\nicefrac{\left(\hat{\sigma}_{i}^{2}-\sigma_{i}^{2}\right)}{\sigma_{i}^{2}}$
using $\sigma_{i}^{2}$ computed exactly using the Takahashi equations.
The maximum error is computed as $\max_{i}\left|r_{i}\right|$ and
the RMSE is computed empirically for $r_{i}$ across all nodes for
each method. For each $\hat{\sigma}_{i}^{2}$ for the block RBMC method,
we also compute a confidence interval (CI) based on the $\chi^{2}$-distribution
in Eq.~(\ref{eq:RBMCestDiagDistr}), with $\sigma_{i}^{2}=\hat{\sigma}_{i|\mathcal{I}^{c}}^{2}$,
and count the share of nodes for which the true value $\sigma_{i}^{2}$
is outside the CI. The CI computation for the MC method uses that
$\hat{\sigma}_{MC,i}^{2}\sim\frac{\sigma_{i}^{2}}{N_{s}}\chi_{N_{s}}^{2}$.
For the Hutchinson method we do not have a simple method to compute
CIs, so this measure is not reported. The lattice is of size $80\times80\times80=512,000$
and for block RBMC we use 5, 10 or 20 blocks in each dimension and
we present results using 20 and 100 random samples.

\begin{table}[H]
\caption{\label{tab:timingsSim}Computing times, empirical errors and proportion
of confidence intervals not covering the true value when computing
the posterior covariance matrix diagonal of a theoretical spatial
model on a $80\times80\times80$ lattice, using different methods
and different number of random samples. The presented results are
averages $\pm$ one standard deviation across 100 runs with different
random seeds. }
\includegraphics{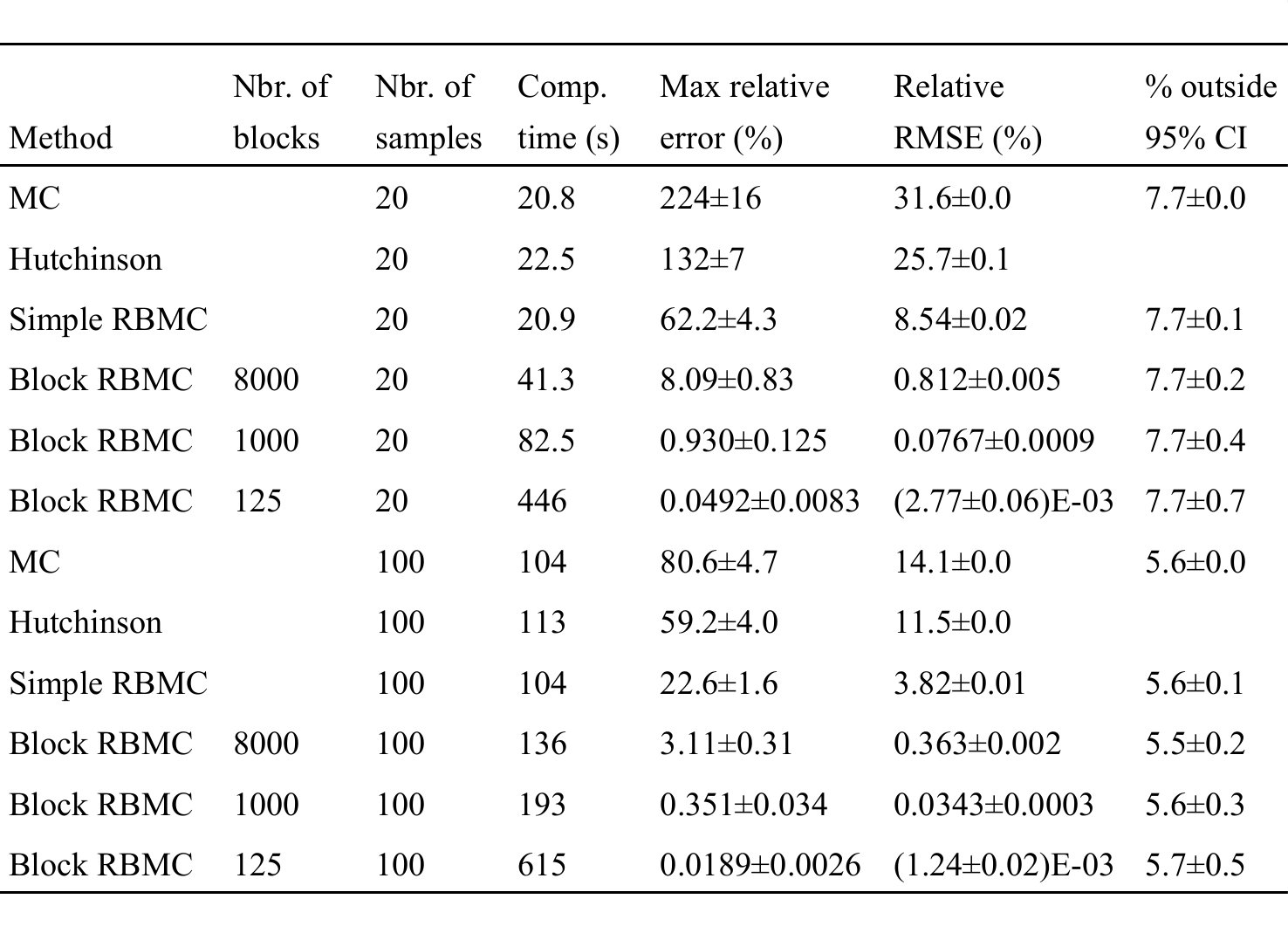}
\end{table}

Table~\ref{tab:timingsSim} clearly shows that our simple RBMC method
performs significantly better than previous methods (MC and Hutchinson)
with the same computing time. We see that the error can be further
decreased using the block RBMC method, but with additional computational
cost. If low error is desirable, the results indicate that block RBMC
with few samples is preferable over simple RBMC with many samples,
as block RBMC with 1000 blocks and 20 samples gives far lower error
than simple RBMC with 100 samples, in less time. Out of the two previous
methods, Hutchinson seems to give lower error than MC, but we noted
a drawback in that Hutchinson can sometimes produce negative variance
estimates. The computed CIs can be seen to cover close to the desired
$95\%$ of the true values, but they are slightly biased because $\hat{\sigma}_{i}^{2}$
is used in place of $\sigma_{i}^{2}$. The bias is reduced when using
a larger number of samples, and since it is so systematic it could
probably be corrected for, knowing the distribution of $\hat{\sigma}_{i}^{2}$.
However, for most applications, this level of error in the uncertainty
of the estimated covariances is likely to be acceptable, so we leave
such a task to future work.

Next, we consider a spatial regression model for neuroimaging \citep{Sid??n2017,Penny2007}.
Brain activity is modeled as a GMRF on a 3D lattice of voxels over
the brain, with $K$ different variables in each voxel, corresponding
to activations of different tasks and an intercept. The resulting
variational Bayes (VB) posterior is a GMRF of size $KN$, where $N$
is the number of voxels, and the Markov assumptions of the model makes
all non-adjacent voxels conditionally independent. This makes the
use of our developed methods possible, if we for the iterative interface
include all $K$ variables in each voxel on for example the frame
to the corresponding interface set $\mathcal{V}_{i}$. We present
results for the block RBMC and iterative interface methods in Table~\ref{tab:timingsfMRI}
for data simulated in the same way as in \citet[Appendix D]{Sid??n2017}
on a $50\times50\times40$ lattice and $K=5$ (the resulting GMRF
has $500,000$ variables). The errors are computed relative to the
exact values computed using the Takahashi equations. We use $N_{s}=20$
samples in $\mathbf{X}$ for both methods and our experience has shown
that the iterative interface methods does not improve much after the
first iteration, so we use $N_{iter}=1$, and we use 5, 10 and 15
interface blocks in each of the three dimensions.

Table~\ref{tab:timingsfMRI} shows that the iterative interface method
can reduce the error beyond what is achievable using the block RBMC
method, but that it requires more computing time and memory. The computing
time of the exact Takahashi equations could probably be significantly
reduced by optimizing the code, but its large memory requirements
(55GB) is just that of storing the Cholesky factor and the corresponding
sparse inverse of $\mathbf{Q}$, which is difficult to reduce further,
showing the infeasibility of exact methods for large problems. The
iterative interface method can also be rather costly memory-wise,
but by choosing the appropriate block sizes one could adapt the memory
usage to the current limitations.

\begin{table}[H]
\caption{\label{tab:timingsfMRI}Computing times, memory usage and errors relative
to the exact values from the Takahashi equations, when computing the
posterior covariance matrix of the spatial model in \citet{Sid??n2017}
using simulated fMRI data on a $50\times50\times40$ with a 5-dimensional
variable in each lattice point, using different methods and different
number of blocks and $20$ random samples. The presented results are
averages $\pm$ one standard deviation across 10\textcolor{blue}{{}
}runs with different random seeds, except for the Takahashi equations.}
\includegraphics{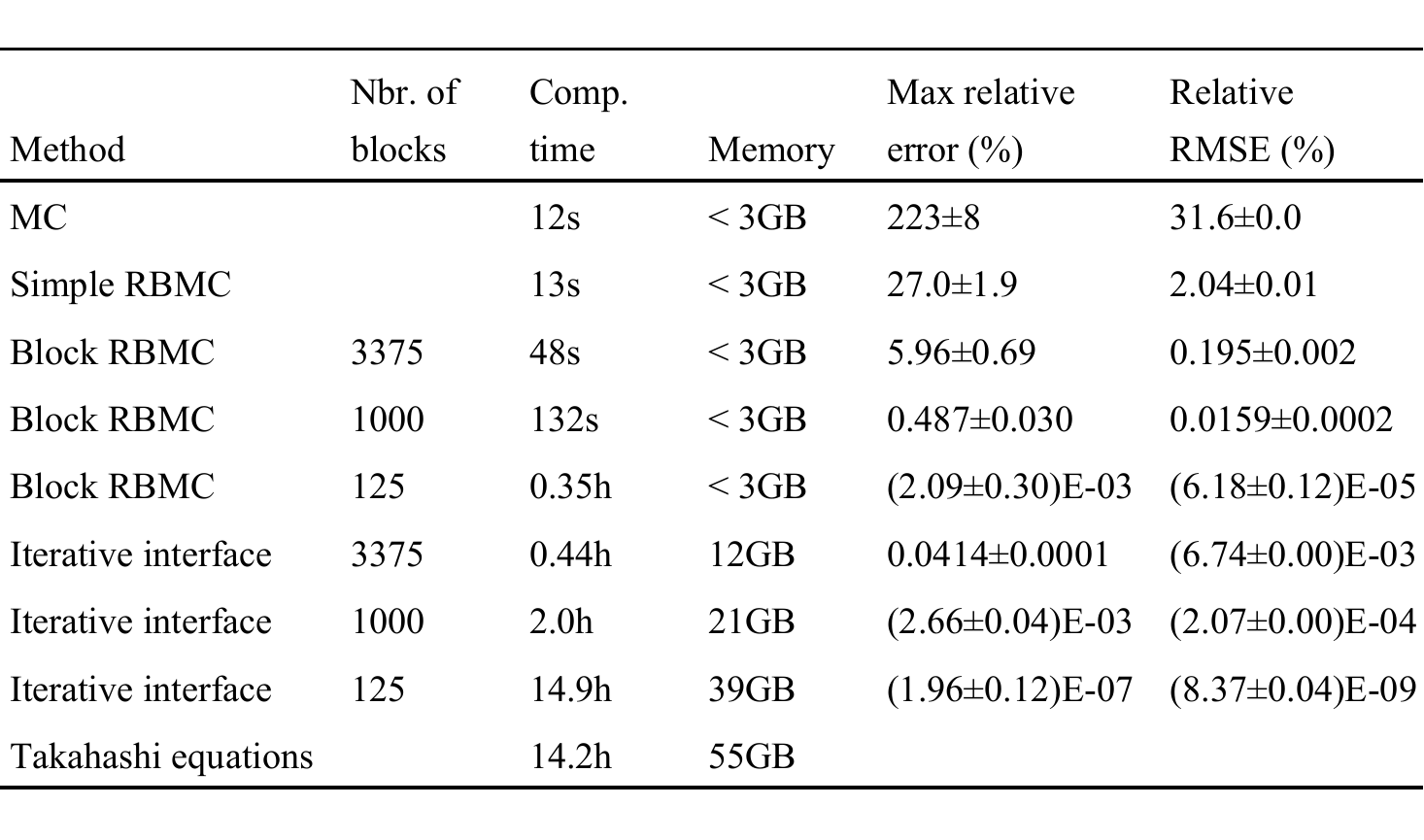}
\end{table}

So far we have only evaluated the methods on marginal variances, but
the methods can be used to estimate all covariances. To show that
the error is small also for the covariances we computed the empirical
absolute RMSE for $\hat{\Sigma}_{i,j}$ across all $i$ and $j$ corresponding
to the same variable in adjacent voxels, estimated with the iterative
interface method with 1000 blocks (the relative RMSE is not suitable
for covariances as they can be zero or negative). The RMSE was $4.04\cdot10^{-5}$
for these off-diagonal elements, which can be compared to $3.02\cdot10^{-5}$
for the diagonal elements, indicating that the errors are in the same
order of magnitude.

\begin{figure}[H]
\includegraphics[width=1\linewidth]{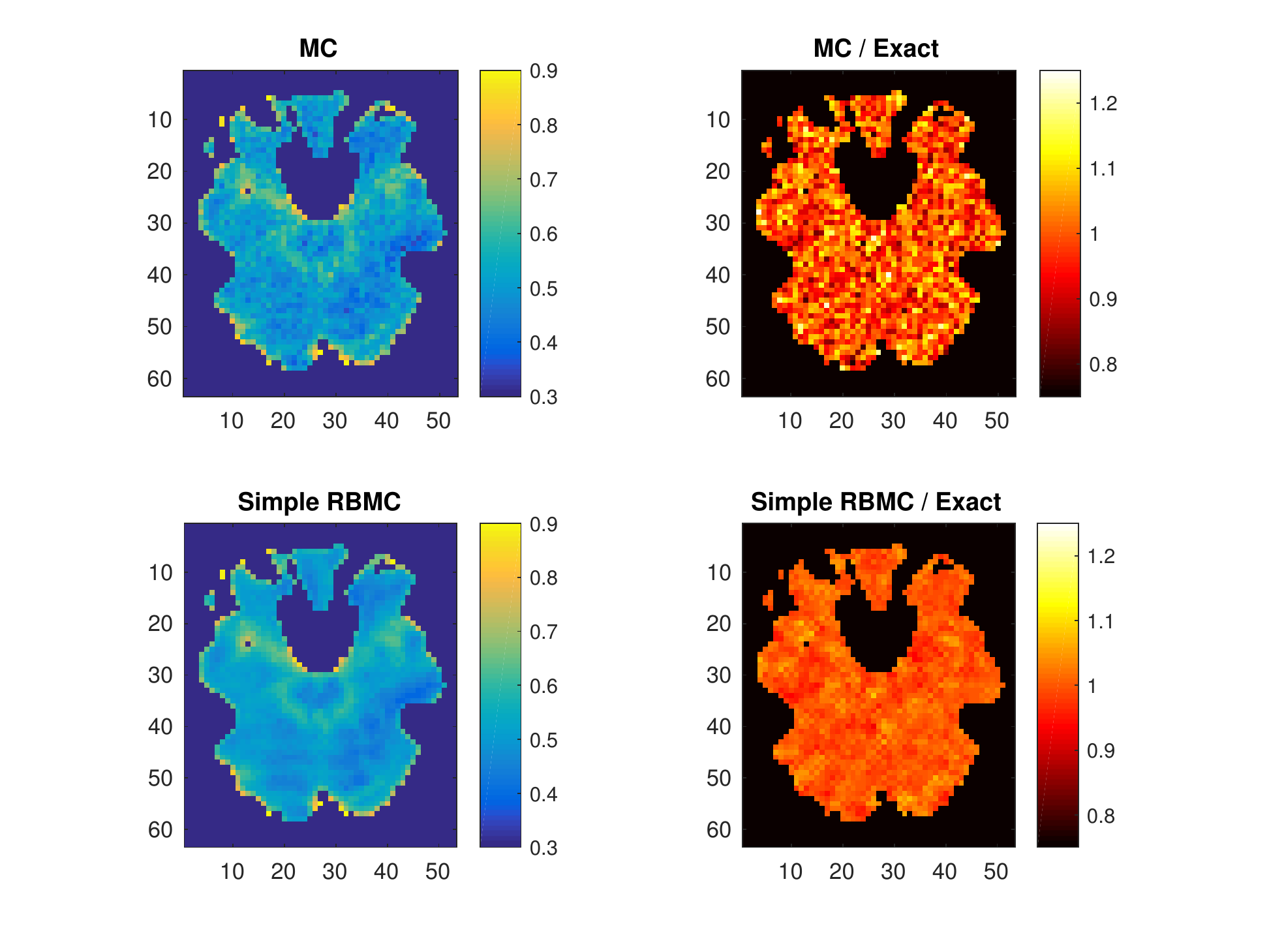}

\caption{Posterior marginal standard deviation estimates for the fMRI data
and model in \citet{Sid??n2017}, based on MC estimation (top row)
and simple RBMC estimation (bottom row). The second column shows the
estimated standard deviations divided with the exact values, computed
using the slower Takahashi equations method. \label{fig:fMRIcomp}}
\end{figure}

Finally, we visualize the improvement of our methods on some real
fMRI data. The top left subfigure in Fig.~\ref{fig:fMRIcomp} replicates
the bottom middle subfigure in \citet[Fig. 3]{Sid??n2017}, showing
MC estimated marginal standard deviations of brain activity over a
brain slice. The top right subfigure shows the ratio ($\hat{\sigma}_{i}/\sigma_{i}$)
compared to the exact marginal standard deviations computed with the
Takahashi equations. The bottom row shows the same, but with simple
RBMC estimates instead of MC. It is clear that the simple RBMC estimates
have much smaller error and using the even more exact block RBMC or
iterative interface methods would reduce the error to levels that
would be hardly visible to the naked eye.

\section{Discussion and future work\label{sec:Discussion}}

The results show that our suggested methods, the simple RBMC, the
block RBMC, and the iterative interface method outperform other sampling-based
methods in terms of accuracy for a given computing time, and exact
methods in terms of memory usage. 

For a practical problem, one could find the desired balance between
error, computing time and memory requirements by choosing between
our proposed methods, the number of samples and the number of blocks
(or block sizes). For the RBMC methods, the error will decrease linearly
with $\sqrt{N_{s}}$, while time and memory requirements grow linearly
with $N_{s}$. As usual with Monte Carlo methods this gives asymptotical
exactness, but this limit is not attainable in practice. 

Both the RBMC and iterative interface methods also converges with
block size, as $N_{b}=1$ will be the same as the exact Takahashi
equations. Exactly how the error, computing time and memory depend
on the block sizes is a difficult question to answer in general, but
by assuming a field that is fairly stationary, the following strategy
could be employed to find the required block size for a given error:
Compute the block RBMC estimates for just one or a small number of
blocks and also compute the corresponding uncertainty measures, as
explained in Section~\ref{subsec:Approximation-variance-and}. Redo
this procedure with increasing block sizes until the uncertainty is
sufficiently small, before computing the estimates for the whole domain.
Since the iterative interface method uses block RBMC for starting
values and then reduces the error, this gives an upper bound for the
uncertainty also for that method. Especially, models with longer spatial
correlation range will require larger blocks to obtain a given accuracy.

There are a number of ways in which our algorithms can be parallelized.
All samples in $\mathbf{X}$ are independent, so these can be generated
in parallel. Also, all steps in the RBMC and iterative interface methods
are done independently for each block and are straightforward to parallelize
over blocks, apart from phase two in the iterative interface method,
but this phase can probably be run in parallel by letting different
threads operate on separate parts of the domain.

As we mentioned in Section~\ref{sec:Background}, our developed methods
can be used for trace estimation needed for EM and VB, but they could
also be used in other methods, for example integrated nested Laplace
approximation (INLA) \citep{Solis-Trapala2009}. INLA normally uses
the Cholesky factor of $\mathbf{Q}$ for computing marginal posterior
variances and $\log\text{\ensuremath{\left|\textbf{Q}\right|}}$.
To avoid the Cholesky factorization, the variances could instead be
approximated using our methods and the log determinant could be approximated
for example using the methods in \citet{Aune2014} or \citet{Ubaru2017}.
The usefulness of our methods within MCMC algorithms is probably limited,
as posterior samples can normally be sampled using $\mathbf{Q}$ directly
with for example the method in \citet{Papandreou2010}, without the
need of computing elements of the covariance matrix. However, our
algorithms could possibly be employed in MCMC post processing to more
efficiently compute marginal variances. The usefulness of our methods
for models not always formulated using precision matrices such as
vector autoregressive (VAR) models \citep{Koop2013} and could also
be further explored.

\section{Conclusions\label{sec:Conclusions}}

We presented a number of methods for estimating selected elements
of the covariance matrix when the precision matrix is sparse and the
corresponding Gaussian density can be sampled from, but too large
for full inversion or even Cholesky factorization. Our methods extends
the idea of \citet{Papandreou2010} to use MC sampling to estimate
covariances, but better utilizes the information from the known precision
matrix to reduce the error, while simultaneously having lower computational
requirements than known exact methods.

\section*{Acknowledgements}

We thank Johan Lindström for the original C++ code implementing the
Takahashi equations. This work was funded by Swedish Research Council
(Vetenskapsrådet) grant no 2013-5229 and grant no 2016-04187. Finn
Lindgren was funded by the European Union's Horizon 2020 Programme
for Research and Innovation, no 640171, EUSTACE.\bibliographystyle{apalike}
\bibliography{/Users/persi28/Dropbox/phd/Artiklar/Bibtex/library,/Users/persi28/Dropbox/phd/Artiklar/Bibtex/manually}

\end{document}